 \documentclass[journal,comsoc]{IEEEtran}
\usepackage[T1]{fontenc}

\usepackage{amsmath}
\usepackage{amsmath,amsthm}
\usepackage{cite}   
\usepackage{graphicx}  
\usepackage{amsmath, amsopn, psfrag, amsthm}   
\usepackage{amssymb}
\usepackage{color}

 \usepackage[caption=false]{subfig} 
\usepackage{algorithm}
\usepackage{amsmath}
\usepackage{algorithmic}
\usepackage[utf8]{inputenc}
\usepackage[english]{babel}
\newtheorem{theorem}{Theorem}
\newtheorem{lemma}{Lemma}
\newtheorem{proposition}{Proposition}
\newtheorem{remark}{Remark}
\usepackage{psfrag}
\usepackage{epsfig}
\graphicspath{{./img/}}
\DeclareGraphicsExtensions{.pdf,.jpeg,.png, .eps}
\DeclareMathOperator*{\argmax}{arg\, max}

\usepackage{amssymb,amsmath,mathtools}

\def\bgama{{\boldsymbol{\Gamma}}}
\def\bPhi{{\boldsymbol{\Phi}}}
\def\bep{{\boldsymbol{\epsilon}}}

\def\bup{{\boldsymbol{\Upsilon}}}
\def\blam{{\boldsymbol{\Lambda}}}
\def\bz{{\boldsymbol{\zeta}}}
\def\bn{{\boldsymbol{\nu}}}
\def\bv{{\boldsymbol{\varrho}}}
\def\bk{{\boldsymbol{\kappa}}}
\def\bdel{{\boldsymbol{\Delta}}}
\def\bpsi{{\boldsymbol{\Psi}}}
\def\bsigma{{\boldsymbol{\sigma}}}

\def\bgama{{\boldsymbol{\Gamma}}}
\def\bPhi{{\boldsymbol{\Phi}}}
\def\bep{{\boldsymbol{\epsilon}}}

\def\bup{{\boldsymbol{\Upsilon}}}
\def\blam{{\boldsymbol{\Lambda}}}
\def\bz{{\boldsymbol{\zeta}}}
\def\bn{{\boldsymbol{\nu}}}
\def\bv{{\boldsymbol{\varrho}}}
\def\bk{{\boldsymbol{\kappa}}}
\def\bdel{{\boldsymbol{\Delta}}}
\usepackage{cite}   
\usepackage{graphicx}  
\usepackage{amsmath, amsopn}   
\usepackage{amssymb}
\usepackage{color}
\usepackage{psfrag}
\usepackage{epsfig}
\usepackage{stfloats}
\usepackage{lipsum}
\usepackage{multicol}
\graphicspath{{./img/}}
\DeclareGraphicsExtensions{.pdf,.jpeg,.png}
\begin{document}
\title{Max-Min Rate of Cell-Free Massive MIMO Uplink with Optimal Uniform Quantization}
\author{Manijeh Bashar,~\IEEEmembership{Student Member,~IEEE}, Kanapathippillai Cumanan,~\IEEEmembership{Member,~IEEE}, Alister G. Burr,~\IEEEmembership{Senior Member,~IEEE}, Hien Quoc Ngo,~\IEEEmembership{Member,~IEEE}, Merouane Debbah,~\IEEEmembership{Fellow,~IEEE}, and Pei Xiao,~\IEEEmembership{Senior Member,~IEEE}
	\thanks{M. Bashar, K. Cumanan and A. G. Burr are with the Department of Electronic Engineering, University of York, Heslington, York, U.K. e-mail: \{mb1465, kanapathippillai.cumanan, alister.burr\}@york.ac.uk. M. Bashar is also with
	 home of the 5G Innovation Centre, Institute for Communication Systems, University of Surrey, U.K. e-mail: m.bashar@surrey.ac.uk. H. Q. Ngo is with the School of Electronics, Electrical Engineering and Computer Science, Queen's University Belfast, Belfast, U.K. e-mail: hien.ngo@qub.ac.uk. M. Debbah is with the Large Networks and Systems Group (LANEAS), CentraleSupelec,
		Universite Paris-Saclay, Gif-sur-Yvette 91192, France, and also with the
		Mathematical and Algorithmic Sciences Lab, Huawei Technologies Co., Ltd.,
		Boulogne-Billancourt 92100, France. e-mail: merouane.debbah@centralesupelec.fr.
		Pei Xiao is with home of the 5G Innovation Centre, Institute for Communication Systems, University of Surrey, U.K. e-mail: p.xiao@surrey.ac.uk.
		}
		\thanks{ The work of K. Cumanan and A. G. Burr was supported by H2020-MSCA-RISE-2015 under grant number 690750.}
		\thanks{The work of H. Q. Ngo was supported by the UK Research and Innovation Future Leaders Fellowships under Grant MR/S017666/1.}
		\thanks{The work of P. Xiao was supported in part by the European Commission under the 5GPPP project 5GXcast
(H2020-ICT-2016-2 call, grant number 761498) as well as by the U.K. Engineering and Physical Sciences  Research  Council  under Grant EP/ R001588/1.
} 
}
\maketitle


\begin{abstract}
Cell-free Massive multiple-input multiple-output (MIMO) is considered, where distributed access points (APs)
multiply the received signal by the conjugate of the estimated channel, and send back a quantized version of this weighted signal to a central processing unit (CPU). For the first time, we present a performance comparison between the case of perfect fronthaul links, the case when the quantized version of the estimated channel and the quantized signal are available at the CPU, and the case when only the quantized weighted signal is available at the CPU. The Bussgang decomposition is used to model the effect of quantization. The max-min problem is studied, where the minimum rate is maximized with the power and fronthaul capacity constraints. To deal with the non-convex problem, the original problem is decomposed into two sub-problems (referred to as receiver filter design and power allocation). Geometric programming (GP) is exploited to solve the power allocation problem whereas a generalized eigenvalue problem is solved to design the receiver filter. An iterative scheme is developed and the optimality of the proposed algorithm is proved through uplink-downlink duality. A user assignment algorithm is proposed which significantly improves the performance. Numerical results demonstrate the superiority of the proposed schemes.
\end{abstract}
\section{Introduction}
Cell-free Massive multiple-input multiple-output (MIMO) has been recognized as a potential technology for 5th Generation (5G) systems, where large number of distributed access points (APs) serve a much smaller number of users, and hence, uniformly good service performance for all users is ensured \cite{
eehienfree,marzetta_free16,ouricc1,ouricc2,our_icc19_noma}. Interestingly, in \cite{marzetta_free16}, it is shown that the system performance of cell-free Massive MIMO depends only on large-scale fading, i.e., the small-scale fading and noise can be averaged out when number of APs is large. In \cite{buzzicellfree} a user-centric approach is proposed where each user is served by a small number of APs. Cell-free Massive MIMO effectively implements a user-centric approach \cite{emil_magazin_ubiquitous}. In \cite{Nanjing_stochstic_tvt18}, the authors consider distributed Massive MIMO in a multi-cell manner, which is different from cell-free massive MIMO (as there is no cell concept).

One of the main issues of cell-free Massive MIMO systems which requires more investigation is the limited-capacity fronthaul links from the APs to a central processing unit (CPU). 
The assumption of infinite fronthaul in \cite{marzetta_free16,eehienfree,alister_pimrc_18}
is not realistic in practice. The fronthaul requirements for Massive MIMO systems, including small-cell and macro-cell base stations (BSs) have been investigated in \cite{backmacro}. The fronthaul load is the main challenge in any distributed antenna systems \cite{backmacro,our_icc19_ee}.
First, we consider the case where all APs send back the quantized version of the minimum mean-square error (MMSE) estimate of the channel from each user and the quantized version of the received signal to the CPU. We next study the case when each AP multiplies the received signal by the conjugate of the estimated channel from each user, and sends back a quantized version of this weighted signal to the CPU. We derive the total number of bits for both cases and show that given the same fronthaul capacity for both cases, the relative performance of the aforementioned cases depends on the number of antennas at each AP, the total number of APs and the channel coherence time.
A new approach is provided to the analysis
of the effect of fronthaul quantization on the uplink of cell-free Massive MIMO.
While there has been significant work in the context of network MIMO on compression techniques such as
Wyner-Ziv coding for interconnection of distributed base stations, here for
simplicity (and hence improved scalability) we assume simple
uniform quantization. We exploit the Bussgang
decomposition \cite{Zillmann} to model the effect of quantization. 

In \cite{marzetta_free16,eehienfree,elinaprecodingfree} the authors propose that the APs design the linear receivers based on the estimated channels, and that this is carried out locally at the APs. Hence, the CPU exploits only the statistics of the channel for data detection. However, in this paper, we propose to exploit a new receiver filter at the CPU to improve the performance of cell-free Massive MIMO systems. The coefficients of the proposed receiver filter are designed based on only the statistics of the channel, which is different from the linear receiver at the APs. The proposed receiver filter provides more freedom in the design parameters and hence, significantly improves the performance of the uplink of cell-free Massive MIMO. The work in \cite{elina_asilomar18} presents a large scale fading decoding (LSFD) postcoding vector and power allocation scheme to solve max-min signal-to-interference-plus-noise ratio (SINR) problem. However, note that the work in \cite{elina_asilomar18} does not present any iterative algorithm to jointly solve power minimization problem and LSFD postcoding vector design. In \cite{our_asilomar18}, the authors use a bisection search approach to solve the power allocation problem. Next, MMSE receiver is exploited to determine the LSFD postcoding vectors. However, in our work, we exploit geometric programming (GP) to optimally solve the power allocation problem. Moreover, we prove that the proposed algorithm is optimal whereas the authors in \cite{elina_asilomar18} does not present any proof of optimality. In addition, the work in \cite{elina_asilomar18} does not consider any quantization errors whereas our work investigates the realistic assumption of limited-capacity fronthaul links.

We next investigate an uplink max-min rate problem with limited fronthaul links. In particular, the receiver filter coefficients and power allocation are optimized in the proposed scheme whereas the work in \cite{marzetta_free16} only considered user power allocations. In particular, we propose a new approach to solve this max-min problem. A similar max–min rate
problem based on SINR known as \textit{SINR balancing} in the
literature has been considered
\cite{cuma_secracy_tvt14,cuma_rate_icc11,cuma_jstsp_16,cuma_jointbf_twc10,cuma_sinr_spl10,rahul_wcnc,cuma_mmse_tvt13,tvt_cuma_Rahulamathavan_mixed}. In \cite{wies06,cai_maxmin_tsp11}, the authors consider MIMO systems and study the problem of max-min user rate to maximize the smallest user rate. The problem of uplink-downlink duality has been investigated in \cite{Bochetvt4,tse}. Note that none of the previous works on uplink-downlink duality consider Massive MIMO and the SINR formula in single-cell does not include any pilot contamination, channel estimation and quantization errors.
To tackle the non-convexity of the original max-min rate problem, we propose to decouple the original problem into two sub-problems, namely, receiver filter coefficient design, and power allocation. We next show that the receiver filter coefficient design problem may be solved through a generalized eigenvalue problem \cite{bookematrix}. Moreover, the user power allocation problem is solved through standard GP \cite{Chiangbook2_gp}. We present an iterative algorithm to alternately solve each sub-problem while one of the design parameters is fixed. Next an uplink-downlink duality for cell-free Massive MIMO system with limited fronthaul links is established to validate the optimality of the proposed scheme. We finally propose an efficient user assignment algorithm and show that further improvement is achieved by the proposed user assignment algorithm. 

The idea of exploiting an iterative algorithm to design the receiver filter and power coefficients in cell-free Massive MIMO system has been proposed in \cite{ourjournal1}. However, in \cite{ourjournal1}, the authors investigate a cell-free Massive MIMO with single-antenna APs and perfect fronthaul links whereas in the this work we exploit a cell-free Massive MIMO system with multiple-antenna APs and limited-capacity fronthaul links. Furthermore, in this work, unlike \cite{ourjournal1}, user assignment is investigated. The contributions of the paper are summarized as follows: 
\begin{itemize}
\item[\textbf{1.}] We consider two cases: i) the quantized versions of the channel estimates and the received signals at the APs are available at the CPU and ii) the quantized versions of processed signals at the APs are available at the CPU. The corresponding achievable rates are derived by using the Use-and-then-Forget (UaF) bounding technique taking into account the effects of channel estimation error and quantization error.

\item[\textbf{2.}] We make use of the Bussgang
decomposition to model the effect of quantization and present the analytical solution to find the optimal step size of the quantizer.

\item[\textbf{3.}] We propose a max-min fairness power control problem which maximizes the smallest of all user rates under the per-user power and fronthaul capacity constraints. To solve this problem, the original problem is decomposed into two sub-problems and an iterative algorithm is developed. The optimality of the proposed algorithm is proved through establishing the uplink-downlink duality for the cell-free Massive MIMO system with limited fronthaul link capacities.

\item[\textbf{4.}] A novel and efficient user assignment algorithm based on the capacity of fronthaul links is proposed which results in significant performance improvement.
\end{itemize}
The rest of the paper is organized as follows. Section II describes
the system model and Section III provides performance analysis. The proposed max-min rate scheme is presented in Section IV and the convergence is provided in Section V. The optimality of the proposed scheme is proved in Section VI. Section VII investigates
the proposed user assignment algorithm. Numerical results are presented in Section VIII, and finally Section IX concludes the paper. 
\begin{figure}
\center
\includegraphics[width=76mm]{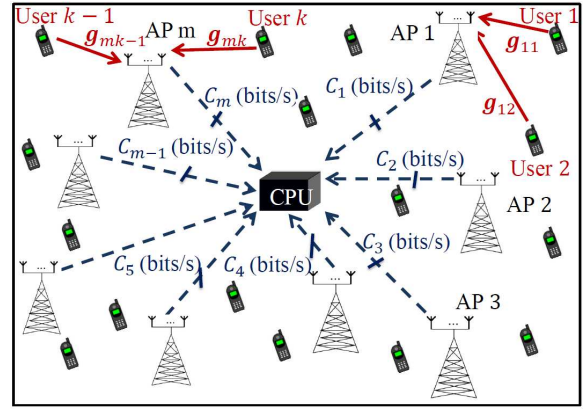}
	\vspace{-.05in}
\caption{The uplink of a cell-free Massive MIMO system with $K$ single-antenna users and \textit{M} APs. Each AP is equipped with $N$ antennas. The solid lines denote the uplink channels and the dashed lines present the limited-capacity fronthaul links from the APs to the CPU.}
\label{sysmodel}
\vspace{-0.5cm}
\end{figure}
\vspace{-.1cm}
\section{SYSTEM MODEL}
We consider uplink transmission in a cell-free Massive MIMO system with $M$ APs and $K$ single-antenna users randomly distributed in a large area. Moreover, we assume each AP has $N$ antennas. The channel coefficient vector between the $k$th user and the $m$th AP, $\mathbf{g}_{mk} \in \mathbb{C}^{N\times 1}$, is modeled as
$
\mathbf{g}_{mk}=\sqrt{\beta_{mk}}\mathbf{h}_{mk},
$
where $\beta_{mk}$ denotes the large-scale fading, the elements of $\mathbf{h}_{mk}$ are independent and identically distributed (i.i.d.) $\mathcal{CN}(0,1)$ random variables, and represents the small-scale fading \cite{marzetta_free16}.
\vspace{-.15cm}
\subsection{Uplink Channel Estimation}
All pilot sequences transmitted by the $K$ users in the channel estimation phase are collected in a matrix $\bPhi \in \mathbb{C}^{\tau_p\times K}$, where $\tau_p$ is the length of the pilot sequence for each user and the $k$th column, $\pmb{\phi}_k$, represents the pilot sequence used for the $k${th} user. After performing a de-spreading operation, the MMSE estimate of the channel coefficient between the $k$th user and the $m$th AP is given by \cite{marzetta_free16}
\vspace{-.04cm}
\begin{IEEEeqnarray}{rCl}
	\small
	\!\!\!\!\!\hat{\mathbf{g}}_{mk}\!= \!c_{mk}\!\left(\!\sqrt{\tau_p p_p}\mathbf{g}_{mk}\!+\!\sqrt{\tau_p p_p}\sum_{k^\prime\ne k}^{K}\mathbf{g}_{mk^\prime}\pmb{\phi}_{k^\prime}^H\pmb{\phi}_{k}\!+\!\mathbf{W}_{p,m}\pmb{\phi}_k\!\right),~
	\label{ghat}
\end{IEEEeqnarray}
where $\mathbf{W}_{p,m} \in \mathbb{C}^{M\times K}$ denotes the noise sequence at the $m$th AP whose elements are i.i.d. $\mathcal{CN}(0,1)$, $p_p$ represents the normalized SNR of each pilot sequence (which we define in Section \ref{sec_numer}), and
$
c _{mk}=\frac{\sqrt{\tau_p p_p}\beta_{mk}}{\tau_p p_p\sum_{k^\prime=1}^{K}\beta_{mk^\prime}|\pmb{\phi}_{k^\prime}^H{\pmb{\phi}}_{k}|^2+1}.
$
Note that, as in \cite{marzetta_free16}, we assume that the large-scale fading,
$\beta_{mk}$, is known.\footnote{
The large-scale fading $\beta_{mk}$ changes very slowly with time. Compared to the small-scale fading, the large-scale fading changes much more slowly, some 40 times slower according to \cite{Rappbook, Ashikhmin_beta}. Therefore, $\beta_{mk}$ can be estimated in advance. One simple way is that the AP takes the average of the power level of the received signal over a long time period. A similar technique for collocated Massive MIMO is discussed  in Section III-D of \cite{Ashikhmin_beta}.} The investigation of cell-free Massive MIMO with realistic COST channel model \cite{ourglobecom_cost,tvt_cost_me,our_ew} will be considered in our future work. 
\subsection{Optimal Quantization Model} 
Based on Bussgang's theorem \cite{Zillmann}, a nonlinear output of a quantizer can be represented as a linear function as follows:
\begin{equation}
\mathcal{Q} (z) = h(z)=az+n_d, ~\forall k, 
\label{zz_zillman}
\end{equation}
where $a$ is a constant value and $n_d$ refers to the distortion noise which is uncorrelated with the input of the quantizer, $z$. The term $a$ is given by 
\begin{equation}
a=\frac{\mathbb{E}\left\{zh(z)\right\}}{\mathbb{E}\{z^2\}}=\frac{1}{p_z}\int_{\mathcal{Z}}zh(z)f_z(z)d~z ,
\label{ee_bussa}
\end{equation}
where $p_z=\mathbb{E}\{|z|^2\}=\mathbb{E}\{z^2\}$ is the power of $z$ and we drop absolute value as $z$ is a real number, and $f_z(z)$ is the probability distribution function of $z$. Denote by\footnote{
Equations (\ref{zz_zillman})-(\ref{b_zillman}) come from \cite{Zillmann} but we include them here for completeness, and to define the terms we used.}
\begin{equation}\label{b_zillman}
b=\frac{\mathbb{E}\left\{h^2(z)\right\}}{\mathbb{E}\{z^2\}}=\frac{1}{p_z}\int_{\mathcal{Z}}h^2(z)f_z(z)d~z.
\end{equation}
Then, the signal-to-distortion noise ratio (SDNR) is
\begin{equation}
\text{SDNR}=\frac{\mathbb{E}\left\{(az)^2\right\}}{\mathbb{E}\{n_d^2\}}=\frac{p_za^2}{p_z\left(b-a^2\right)}=\frac{a^2}{b-a^2},
\end{equation}
According to \cite{Zillmann,ourvtc18,dick_tvt_19}, the midrise uniform quantizer function $h(z)$ is given by
\vspace{-.2cm}
\begin{equation}\label{ee_busshz}
\small
h(z) =
\begin{cases}
-\frac{L-1}{2}\Delta & z \le -\left(\frac{L}{2}+1\right)
\Delta,\\
\left(l+\frac{1}{2}\right)\Delta &  l \Delta\le z \le (l+1)\Delta , l=-\frac{L}{2}+1,\cdots,\frac{L}{2}-2,\\
\frac{L-1}{2}\Delta &  z \ge  \left(\frac{L}{2}-1\right)\Delta,
\end{cases}
\end{equation}
where $\Delta$ is the step size of the quantizer and $L=2^\alpha$, where $\alpha$ is number of quantization bits.
\vspace{-.15cm}
\begin{lemma}\label{lemmaab}
The terms $a$ and $b$ are obtained as follows:
	\begin{eqnarray}
 	a\!=\!\Delta\sqrt{\dfrac{2}{\pi p_z}}\!\left(\!\sum_{l=1}^{\frac{L}{2}-1}\!e^{-\dfrac{l^2\Delta^2}{2p_z}}\!+\!1\!\right)\!,
	b\!=\!\dfrac{\Delta^2}{p_z}\!\left(\! \frac{1}{4}\!+\!4\sum_{l=1}^{\frac{L}{2}-1}\!lQ\left(\frac{l\Delta}{\sqrt{p_z}}\right) \right),\!\label{lemmaab2}
	\end{eqnarray}
where $Q(x)$ is the Q-function and is given by $Q(x)=\frac{1}{2}\text{erfc}\left(\frac{x}{\sqrt{2}}\right)$, where erfc refers to the  complementary error function \cite{erfcbook}.
\end{lemma}
{\textit{Proof:}} Please refer to Appendix A.~~~~~~~~~~~~ ~~~~~~~~~~~ $\blacksquare$
\\\
In general, terms $a$ and $b$ are functions of the power of the quantizer input, $p_z$. To remove this dependency, we normalize the input signal by dividing the input signal, $z$, by the square root of its power, $\sqrt{p_z}$, and then multiply the quantizer output by its square root, $\sqrt{p_z}$. Hence, by introducing a new variable $\dot{z}=\frac{z}{\sqrt{p_z}}$, we have
\vspace{-.061cm}
\begin{equation}
\mathcal{Q}(z)=\sqrt{p_z}\mathcal{Q}(\dot{z})
=
\dot{a}
\sqrt{p_z}\dot{z}+\sqrt{p_z}\dot{n}_d
=
\dot{a}z+\sqrt{p_z}\dot{n}_d.
\label{sigmamultiplyq}
\end{equation}
Note that (\ref{sigmamultiplyq}) enables us to find the optimum step size of the quantizer and the corresponding $\dot{a}$. Note that for the case of 
$\dot{\Delta}=\frac{1}{\sqrt{p_z}}\Delta$, we have 
$\dot{a}=a$, 
$\dot{b}=b$.
The optimal step size of the quantizer is obtained by solving the following maximization problem:
\vspace{-.1cm}
{\begin{IEEEeqnarray}{rCl}\label{maxdelat}
\vspace{-.15cm}
&&\!\!\!\!\!\!\!\!\Delta_\text{opt}=\argmax_{\Delta}{\text{SDNR}}\!=\!\argmax_{\Delta}~\frac{{a}^2}{{b}-{a}^2}
\stackrel{I_1}{=}
\argmax_{\dot{\Delta}}~\frac{\dot{a}^2}{
\dot{b}-\dot{a}^2}
\nonumber 
\\
&=&
\argmax_{\dot{\Delta}}\frac{\dot{a}^2}{\dot{b}}
\stackrel{I_2}{\triangleq}
\argmax_{\dot{\Delta}}\left(
\frac{\dfrac{2\dot{\Delta}^2}{\pi}
\left(\sum_{l=1}^{\frac{L}{2}-1}
\exp\left(\frac{-l^2\dot{\Delta}^2}{2}\right)+1\right)^2}
{\dot{\Delta}^2\left(\frac{1}{4}+4\sum_{l=1}^{\frac{L}{2}-1}lQ\left(l\dot{\Delta}\right)\right)}
\right)
\nonumber 
\\
&=&
\argmax_{\dot{\Delta}}\left(\frac{\left(\sum_{l=1}^{\frac{L}{2}-1}2
\exp{\left(-\dfrac{l^2\dot{\Delta}^2}{2}\right)}+1\right)^2}{ \frac{1}{4}
+
4\sum_{l=1}^{\frac{L}{2}-1}l~Q\left(l\dot{\Delta}\right)}\right),
\end{IEEEeqnarray}}where in step $I_1$, we have used (\ref{sigmamultiplyq}) and step $I_2$ comes from results in Lemma 1. Moreover, note that $\dot{\Delta}=\frac{\Delta}{\sqrt{p_z}}$. The maximization problem in (\ref{maxdelat}) can be solved through a one-dimensional search over $\dot{\Delta}$ for a given $L$ in a symbolic mathematics tool such as Mathematica. 
For the input $\dot{z}$ with $p_{\dot{z}}=1$, the optimal step size of the quantizer $\dot{\Delta}_{\text{opt}}$, the resulting distortion noise power, $p_{\dot{n}_d}=\mathbb{E}\{|\dot{n}_d|^2\}=\dot{b}-\dot{a}^2$, and the resulting 
$\dot{a}$ are summarized in Table \ref{tablezillmann}.
\vspace{-.1cm}
\begin{remark}
Interestingly, the optimal values for quantization step size, 
$\dot{\Delta}_{opt}$, given in Table \ref{tablezillmann}, are exactly the same as \textcolor{black}{the optimal values of quantization step size in \cite{max_quantization}. In \cite{max_quantization}, J. Max did not provide any analytical solution to solve the problem of minimizing the mean-squared distortion (or mean-squared error (MSE)) and to obtain the optimal quantization step size.} Moreover, J. Max only calculates the optimal step size and the resulting distortion power for $\alpha=1,\cdots,5$ whereas Lemma \ref{lemmaab} enables us to calculate the optimal step size and the resulting distortion power for any quantization resolution. Values for $\alpha$ up to 10 are listed in Table \ref{tabledelta}.
\end{remark}
\begin{table}[!t]	
\centering 
\caption{The optimal step size and distortion power of a uniform quantizer with Bussgang decomposition.} 
	\vspace{-.04in}
\label{tabledelta} \label{tablezillmann}
 \begin{tabular}{ c c c c}
 \hline\\
  \vspace{-.02cm}
$\alpha$  & 
$\dot{\Delta}_{\text{opt}}$ & $\sigma_{\dot{e}}^2=p_{\dot{n}_d}=\textcolor{black}{\dot{b}}-\textcolor{black}{\dot{a}}^2$ & $\textcolor{black}{\dot{a}}$   
\\ [.1ex] 
 \hline\hline
 \vspace{.01cm}
{1} & 1.596 &  0.2313 &  0.6366\\[.02ex] 
 \hline
 \vspace{.01cm}
{2} &0.9957 &  0.10472 & 0.88115\\ [.02ex] 
 \hline
  \vspace{.01cm}
{3} &0.586 &  0.036037  &0.96256\\ [.02ex] 
 \hline
  \vspace{.01cm}
{4} &0.3352 & 0.011409  & 0.98845\\ [.02ex] 
 \hline
  \vspace{.01cm}
{5} &0.1881  & 0.003482  & 0.996505\\ [.02ex] 
 \hline	
   \vspace{.01cm}
{6} &0.1041  & 0.0010389  & 0.99896\\ [.02ex] 
 \hline	
   \vspace{.01cm}
{7} &0.0568 & 0.0003042  & 0.99969\\ [.02ex] 
 \hline	
   \vspace{.01cm}
{8} &0.0307  & 0.0000876  & 0.999912\\ [.02ex] 
 \hline	
\\
 \end{tabular}
 \vspace{-.6cm}
\end{table}
\vspace{-.1cm}
\subsection{Uplink Transmission}
In this subsection, we consider the uplink data transmission, where all users send their signals to the APs.
The transmitted signal from the $k$th user is represented by
$
x_k= \sqrt{\textcolor{black}{\rho}q_k}s_k,
$
where $s_k$ ($\mathbb{E}\{|s_{k}|^2\} = 1$) and $q_k$ denotes the transmitted symbol and the transmit power from the $k$th user, respectively, where $\rho$ represents the normalized uplink SNR (see Section VIII for more details).
The $N\times 1$ received signal at the $m$th AP from all users is given by 
\vspace{-.01in}
\begin{equation}
\mathbf{y}_m= \sqrt{\rho}\sum_{k=1}^{K}\mathbf{g}_{mk}\sqrt{q_k}s_k+\mathbf{n}_m, 
\label{ym}
\vspace{-.1in}
\end{equation}
where each element of $\mathbf{n}_m \in \mathbb{C}^{N\times 1}$, $n_{n,m}\sim \mathcal{CN}(0,1)$ is the noise at the $m$th AP.
\section{Performance Analysis}
In this section, the performance analysis for two cases is presented. First we consider the case when the quantized versions of the channel estimates and the received signals are available at the CPU. Next, it is assumed that only the quantized versions of the weighted signals are available at the CPU. The corresponding achievable rates are derived by exploiting the UaF bounding technique.

{\textbf{Case 1.} \textit{Quantized Estimate of the Channel and Quantized Signal Available at the CPU}:}
The $m$th AP quantizes the terms $\hat{\mathbf{g}}_{mk}$, $\forall k$, and $\mathbf{y}_m$, and forwards the quantized channel state information (CSI) and the quantized signals in each symbol duration to the CPU. The quantized signal can be obtained as:
\begin{equation}\label{y_quantize}
\mathcal{Q}\left([\mathbf{y}_{m}]_n\right)
=
 \textcolor{black}{\dot{a}}[\mathbf{y}_{m}]_n+[{\mathbf{e}}_{m}^y]_n
=
[\bz_{m}]_n+j[\bn_{m}]_n, ~\forall m ,n, 
\end{equation}
where $[\mathbf{e}_{m}^y]_n$ refers to the quantization error, and $[\bz_{m}]_n$ and $[\bn_{m}]_n$ are the real and imaginary parts of the output of the quantizer, respectively. Note that we separately quantize the imaginary and real parts of the input of the quantizer. \textcolor{black}{Note that $[\mathbf{x}]_n$ represents the $n$th element of vector $\bf{x}$}. The analog-to-digital converter (ADC) quantizes the real and imaginary parts of $[\mathbf{y}_m]_n$ with $\alpha$ bits each, which introduces quantization errors $[\mathbf{e}_{m}^y]_n$ to the received signals \cite{Oppenheimsignal,mythesis}. In addition, the ADC quantizes the MMSE estimate of CSI as:
\begin{equation}\label{g_quantize}
\mathcal{Q}\left([\hat{\mathbf{g}}_{mk}]_n\right)
=
\dot{a}[\hat{\mathbf{g}}_{mk}]_n
+
[{\mathbf{e}}_{mk}^g]_n=[\bv_{mk}]_n+j[\bk_{mk}]_n,\forall k, n, 
\end{equation}
where $[\bv_{mk}]_n$ and $[\bk_{mk}]_n$ denote the real and imaginary parts of the output of the quantizer, respectively. Again, note that the real and imaginary parts of the input of the quantizer are separately quantized.
For simplicity, we assume all APs use the same number of bits to quantize the received signal, $\mathbf{y}_m$, and the estimated channel, $\hat{\mathbf{g}}_{mk}$.
Therefore, $[{\mathbf{e}}_{m}^y]_n
=
\mathbb{E}\{|[\mathbf{y}_{m}]_n|^2\}
 [\dot{\mathbf{e}}_{m}^y]_n$
and 
$[{\mathbf{e}}_{mk}^g]_n
=
\mathbb{E}\{|[\hat{\mathbf{g}}_{mk}]_n|^2\} 
[\dot{\mathbf{e}}_{mk}^g]_n$, where
$\mathbb{E}\left\{\left|[\dot{\mathbf{e}}_{m}^y]_n\right|^2\!\right\}=\mathbb{E}\left\{\left|[\dot{\mathbf{e}}_{mk}^g]_n\right|^2\!\right\}=\sigma_{\dot{e}}^2$. Note that 
$\mathbb{E}\left\{\left|[\dot{\mathbf{e}}_{m}^y]_n\right|^2\!\right\}$ and $\mathbb{E}\left\{\left|[\dot{\mathbf{e}}_{mk}^g]_n\right|^2\!\right\}$ are quantization errors of a quantizer with normalized input 
$[\dot{\mathbf{y}}_{m}]_n = \frac{[\mathbf{y}_{m}]_n}{\sqrt{\mathbb{E}\{|[\mathbf{y}_{m}]_n|^2\}}}$ 
and 
$[\dot{\hat{\mathbf{g}}}_{mk}]_n=\frac{[\hat{\mathbf{g}}_{mk}]_n}{\sqrt{\mathbb{E}\{|[\hat{\mathbf{g}}_{mk}]_n|^2\}}}$, respectively. Note that due to power normalization, $\textcolor{black}{\dot{a}}$, $\textcolor{black}{\dot{b}}$, and optimal step size for (\ref{y_quantize}) and (\ref{g_quantize}) are the same and provided in Table \ref{tablezillmann}.
The received signal for the $k$th user after using the maximum ratio combining (MRC) detector at
the CPU is given by
\vspace{-.02in}
\begin{IEEEeqnarray}{rCl} \label{rkcase1}
	&&\!\!\!\!\!\!\!r_k
	 \!\!=\!\!
	 \sum_{m=1}^{M}\!u_{mk}\!\left(\mathcal{Q}\left(\hat{\mathbf{g}}_{mk}\right)\right)^H\!\mathcal{Q}\left(\mathbf{y}_{m}\right)
	\!=\!\!
	\sum_{m=1}^M\!u_{mk}\!\left(\!\dot{a}\hat{\mathbf{g}}_{mk} 
	\!+\!
	\mathbf{e}_{mk}^{\hat{g}}\right)^H\!\!\left(\textcolor{black}{\dot{a}}\mathbf{y}_m
	\!+\!
	\mathbf{e}_{m}^y\right)
	\nonumber
	\\
	&\!\!=\!\!&
	\sum_{m=1}^M\!u_{mk}\!\left(\!\textcolor{black}{\dot{a}}\hat{\mathbf{g}}_{mk}\!+\!\mathbf{e}_{mk}^{\hat{g}}\right)^H\!
	\left(\!\textcolor{black}{\dot{a}}\sqrt{\rho}\!\sum_{k=1}^{K}\!\mathbf{g}_{mk}\sqrt{q_k}s_k\!+\!\textcolor{black}{\dot{a}}\mathbf{n}_m\!+\!\mathbf{e}_{m}^y\!\right)\!\!
\nonumber
\\
&=&
\textcolor{black}{\dot{a}}^2\underbrace{\sqrt{\rho}\mathbb{E}\left\{\sum_{m=1}^M
u_{mk}\hat{\mathbf{g}}_{mk}^H
		{\mathbf{g}}_{mk}\sqrt{q_k}\right\}}_{\text{DS}_k}s_k+
	\textcolor{black}{\dot{a}}^2\underbrace{\sum_{m=1}^{M}u_{mk}\hat{\mathbf{g}}_{mk}^H\mathbf{n}_m}_{\text{TN}_k}
	\nonumber
	\\&+&\!
	\textcolor{black}{\dot{a}}^2\!\!\underbrace{\sqrt{\rho}\!\left(\!\sum_{m=1}^M
	u_{mk}\hat{\mathbf{g}}_{mk}^H\!{\mathbf{g}}_{mk}\sqrt{q_k}\!-\!\mathbb{E}\!\left\{\!\sum_{m=1}^Mu_{mk}\hat{\mathbf{g}}_{mk}^H{\mathbf{g}}_{mk}\sqrt{q_k}\!\right\}\right)}_{{\text{BU}_k}}\!s_k\!\!+\!\dot{a}^2\!
	\nonumber\\
	&&
	\!\!\!\!\!\!
	\!\!\sum_{k^{\prime}\neq k}^{K}\!\!\underbrace{\sqrt{\rho}\!\sum_{m=1}^M\!\!u_{mk}\hat{\mathbf{g}}_{mk}^H
		{\mathbf{g}}_{mk^\prime}\sqrt{q_{k^\prime}}}_{{\text{IUI}_{kk^\prime}}}\!s_{k^{\prime}}
\!+\!\!\!
\sum_{k^{\prime}= 1}^{K}\!\!\dot{a}\!\underbrace{\sqrt{\rho}\!\!\sum_{m=1}^{M}\!\!u_{mk}(\mathbf{e}_{mk}^{\hat{g}}\!)^H\!{\mathbf{g}}_{mk^\prime}\sqrt{q_{k^\prime}}
		}_{\text{TQE}_{kk^\prime}}s_{k^{\prime}}
		\nonumber
		\\
	&+&\textcolor{black}{\dot{a}}\underbrace{\!\sum_{m=1}^{M}u_{mk}({\mathbf{e}}_{mk}^{g}\!)^H\!\mathbf{n}_m}_{\text{TQE}_k^g}\!+
	\textcolor{black}{\dot{a}}\underbrace{\!\sum_{m=1}^{M}\!u_{mk}\hat{\mathbf{g}}_{mk}^H{\mathbf{e}}_m^y}_{\text{TQE}_k^y}\!+\!
	\underbrace{\!\sum_{m=1}^{M}\!u_{mk}\left(\mathbf{e}_{mk}^{\hat{g}}\!\right)^H\!{\mathbf{e}}_m^y}_{\text{TQE}_k^{gy}},\!\!
\end{IEEEeqnarray}
where $\text{DS}_k$ and $\text{BU}_k$ denote the desired signal (DS) and beamforming uncertainty (BU) for the $k$th user, respectively, and $\text{IUI}_k$ represents the inter-user-interference (IUI) caused by the $k^\prime$th user. In addition, $\text{TN}_k$ accounts for the total noise (TN) following the MRC detection, and finally the terms $\text{TQE}_k^\text{y}$, $\text{TQE}_k^\text{g}$, $\text{TQE}_k^{\text{gy}}$ and $\text{TQE}_{kk^\prime}$ refer to the total quantization error (TQE) at the $k$th user due to the quantization errors at the channel and signal.
Moreover, by collecting all the coefficients $u_{mk}, \forall m$, corresponding to the $k$th user, we define $\mathbf{u}_k = [u_{1k}, u_{2k},\cdots, u_{Mk}]^T$ and without loss of generality, it is assumed that $|| \mathbf{u}_k||=1$. The optimal values of $u_{mk}$ are investigated in Section IV.
\begin{proposition}\label{prop_mutual_uncor}\!Terms $\text{DS}_k$, $\text{BU}_k$, $\text{IUI}_{kk^\prime}$, $\text{TQN}_{kk^\prime}$, $\text{TQN}_k^g$, $\text{TQN}_k^y$, $\text{TQN}_k^{gy}$ are mutually uncorrelated.
\end{proposition}
{\textit{Proof:}} Please refer to Appendix B. ~~~~~~~~~~~~~~~~~~~~~~~~~~~~~~~~$\blacksquare$
\\\\\
To obtain an achievable rate, we use the UaF bounding technique as in \cite{marzetta_free16}. This techniques is commonly used in massive MIMO \cite{hienbook,emil_multislop} since it yields a simple and tight achievable rate which enables us to further design the systems. The tightness of this bound for cell-free Massive MIMO is presented in \cite{marzetta_free16}. Using Proposition 1 and the UaF bounding technique in \cite{marzetta_free16}, 
we can obtain an achievable rate as
$R_k^{\text{Case 1}} = \log_2(1 + \text{SINR}_k^{\text{Case 1}})$, where $\text{SINR}_k^{\text{Case 1}}$ is given by (\ref{sinrk_apx_quan}). The closed-form expression for the achievable uplink rate of the \textit{k}th user is given in the following theorem. 
\begin{small}
\begin{figure*}[t!]
\small
	\begin{IEEEeqnarray}{rCl}
		\hrulefill
		\small
		\label{sinrk_apx_quan}
\!\!\!\!\!\!\!\!\!\text{SINR}_k^{\text{Case 1}}\!\!=
\!\!
\small
\dfrac{\textcolor{black}{\dot{a}}^4\left|\text{DS}_k\right|^2}
		{\textcolor{black}{\dot{a}}^4\mathbb{E}\!\left\{\left|\text{BU}_k\right|^2
		\right\}\!+\!\textcolor{black}{\dot{a}}^4\mathbb{E}\!\left\{\left|\text{TN}_k\right|^2\right\}\!+\!
			\textcolor{black}{\dot{a}}^4\sum_{k^\prime\ne k}^K\mathbb{E}\left\{\left|\text{IUI}_{kk^\prime}\right|^2\!\right\}\!+\!
			\!\textcolor{black}{\dot{a}}^2\mathbb{E}\!\left\{\left|\text{TQE}_k^\text{y}\right|^2\right\}\!+\!\textcolor{black}{\dot{a}}^2\mathbb{E}\left\{\left|\text{TQE}_k^\text{g}\right|^2\right\}\!+\!\!
			\textcolor{black}{\dot{a}}^2\!\!\!\sum\limits_{k^\prime=1}^K
			\!\mathbb{E}\left\{\left|\text{TQE}_{kk^\prime}\right|^2\right\}\!+\!
			\!\mathbb{E}\left\{\left|\text{TQE}_k^\text{gy}\right|^2\right\}}\!.\!
	\end{IEEEeqnarray}
	\begin{IEEEeqnarray}{rCl}
		\label{sinr_er}
\small
\!\!\!\!\!\!\!\!\text{SINR}_k^{\text{Case 1}}
\!\!=\!\!
\dfrac{N^2q_k\left(\sum_{m=1}^{M}u_{mk}\gamma_{mk}\right)^2}
{N^2\sum_{k^\prime\ne k}^K\!q_{k^\prime}\left(\sum_{m=1}^{M}u_{mk}\gamma_{mk}\dfrac{\beta_{mk^\prime}}{\beta_{mk}}\right)^2\left|\pmb{\phi}_k^H\pmb{\phi}_{k^\prime}\right|^2+\!N\left(\dfrac{C_{\text{tot}}}{\textcolor{black}{\dot{a}}^4}+1\right)\sum_{m=1}^{M}u_{mk}\gamma_{mk}\sum_{k^\prime=1}^{K}q_{k^\prime}\beta_{mk^\prime}+\dfrac{N}{\rho}\left(\dfrac{C_{\text{tot}}}{\textcolor{black}{\dot{a}}^4}+1\right)\sum_{m=1}^{M}u_{mk}\gamma_{mk}}\!.~
\vspace{-.01cm}
	\end{IEEEeqnarray}
		\vspace{-.06cm}
		\hrulefill
\end{figure*}
\end{small}			
\begin{theorem}
	\label{theorem_up_quan_case2}
	Having the quantized CSI and the quantized signal at the CPU and employing MRC detection at the CPU, the closed-form expression for the achievable rate of the $k$th user is 
given by $R_k^{\text{Case 1}}=\log_2(1+\text{SINR}_k^{\text{Case 1}})$, where the $\text{SINR}_k^{\text{Case 1}}$ is given by (\ref{sinr_er}) (defined at the top of this page), where $\gamma_{mk}=\sqrt{\tau_p p_p}\beta_{mk}c_{mk}$ and $C_{\text{tot}}=2\textcolor{black}{\dot{a}}^2\sigma_{\dot{e}}^2+\sigma_{\dot{e}}^4$. 
\end{theorem}
\vspace{-.062cm}
{\textit{Proof:}}
The power of quantization errors can be obtained as
\vspace{-.09in}
	\begin{eqnarray}
	&&\mathbb{E}\left\{\left|[\mathbf{e}_{m}^y]_n\right|^2\!\right\}=\mathbb{E}\left\{\left|[\dot{\mathbf{e}}_{m}^y
	]_n\right|^2\!\right\}\left(\rho\sum_{k^\prime=1}^{K}q_{k^\prime}\beta_{mk^\prime}+1\right),
	 \nonumber
	\\
&&	\mathbb{E}\left\{\left|[\mathbf{e}_{mk}^g]_n\right|^2\!\right\}=\mathbb{E}\left\{\left|[\dot{\mathbf{e}}_{mk}^g
	]_n\right|^2\!\right\}\gamma_{mk}\label{quanerr32}.
	\end{eqnarray}
Since $\mathbb{E}\left\{\left|[\dot{\mathbf{e}}_{m}^y]_n\right|^2\!\right\}=\mathbb{E}\left\{\left|[\dot{\mathbf{e}}_{mk}^g]_n\right|^2\!\right\}=\sigma_{\dot{e}}^2$, we have:
\vspace{-.02cm}
	\begin{eqnarray}
	&&\mathbb{E}\left\{\left|[\mathbf{e}_{m}^y]_n\right|^2\!\right\}=\sigma_{\dot{e}}^2\left(\rho\sum_{k^\prime=1}^{K}q_{k^\prime}\beta_{mk^\prime}+1\right),	 \nonumber
	\\
&&
	\mathbb{E}\left\{\left|[\mathbf{e}_{mk}^g]_n\right|^2\!\right\}=\sigma_{\dot{e}}^2\gamma_{mk}\label{samequanerr32}.
	\end{eqnarray}
Using (\ref{quanerr32}) and the fact that quantization error is indepnedent with the input of the quantizer, after some mathematical manipulations, we have:
\vspace{-.03cm}
\begin{IEEEeqnarray}{rCl}\label{tqe_all_case1}
	\!\!\!\!\!\!\!\!&&\!\!\!\!\textcolor{black}{\dot{a}}^2\mathbb{E}  \left\{\left|\text{TQE}_k^\text{y}\right|^2\right\}+
	\textcolor{black}{\dot{a}}^2\mathbb{E}  \left\{\left|\text{TQE}_k^\text{g}\right|^2\right\}+
	\textcolor{black}{\dot{a}}^2\sum_{k^\prime=1}^K
	\!\mathbb{E}\left\{\left|\text{TQE}_{kk^\prime}\right|^2\right\}
	\nonumber\\
	&+&\mathbb{E}  \left\{\left|\text{TQE}_k^\text{gy}\right|^2\right\}
=
	NC_{\text{tot}}\sum_{m=1}^{M}u_{mk}\gamma_{mk}\left(\rho\sum_{k^\prime=1}^{K}q_{k^\prime}\beta_{mk^\prime}+1\right)\!.~~~~
\end{IEEEeqnarray}
Note that the terms $\left|\text{DS}_k\right|^2$, $\mathbb{E}\!\left\{\!\left|\text{BU}_k\right|^2\!\right\}$, and $\mathbb{E}\!\left\{\left|\text{IUI}_{kk^\prime}\!\right|^2\!\right\}$ are derived in (\ref{dsk_vector}), (\ref{ebuk}) and (\ref{euiu_final}), respectively. Finally substituting (\ref{tqe_all_case1}), (\ref{dsk_vector}), (\ref{ebuk}) and (\ref{euiu_final}) into (\ref{sinrk_apx_quan}) results in (\ref{sinr_er}), which completes the proof of Theorem \ref{theorem_up_quan_case2}.~~~~~~~~~~~~~~~~~~~~~~~~~~~~~
~~~~~~~~~~~~~~~~~~~~~~~~$\blacksquare$

{\textbf{Case 2.} \textit{Quantized Weighted Signal Available at the CPU:}}
The $m$th AP quantizes the terms $z_{m,k}= \hat{\mathbf{g}}_{mk}^{H}\mathbf{y}_m$, $\forall k$, and forwards the quantized signals in each symbol duration to the CPU as
\begin{IEEEeqnarray}{rCl}
z_{mk}= \hat{\mathbf{g}}_{mk}^{H}\mathbf{y}_m =r_{mk}+js_{mk}, ~\forall k,m, 
\end{IEEEeqnarray}
where $r_{mk}$ and $s_{mk}$ represent the real and imaginary parts of $z_{mk}$, respectively. An ADC quantizes the real and imaginary parts of $z_{m,k}$ with $\alpha$ bits each, which introduces quantization errors to the received signals \cite{Oppenheimsignal}.
Let us consider the term $e_{mk}^z$ as the quantization error of the $m$th AP. Hence, using the Bussgang decomposition, the relation between $z_{mk}$ and its quantized version, $\dot{z}_{mk}$, can be written as
\vspace{-.01cm}
\begin{IEEEeqnarray}{rCl}
\mathcal{Q}\left(z_{mk}\right) = \textcolor{black}{\dot{a}}z_{mk}+e_{mk}^z. \label{zz}
\end{IEEEeqnarray}
Note that given the fact that the input of quantizer, i.e., $z_{mk}=\hat{\mathbf{g}}_{mk}^{H}\mathbf{y}_m$, is the summation of many terms, it can be approximated as a Gaussian random variable. This enables us to exploit the values given in Table \ref{tabledelta}, which are obtained for Gaussian input. The aggregated received signal at the CPU can be written as\footnote{Note that for both Case 1 and Case 2, the AP estimates the channel. Therefore the total complexity is the same for Case 1 and Case 2. However, in Case 1 the CPU performs $N^2$ multiplication and $M-1$ additions whereas in Case 2 the APs establish $N$ multiplications and the CPU performs $N$ multiplications and combines the transmitted signals form $M$ APs (via fronthaul links) which requires $M-1$ additions.}
\begin{IEEEeqnarray}{rCl}
&&\!\!r_k
\!=\!
\sum_{m=1}^{M}\!u_{mk}\!\Bigg(\!\textcolor{black}{\dot{a}}\underbrace{\hat{\mathbf{g}}_{mk}^H\!\mathbf{y}_{m}}_{z_{mk}}+e_{mk}^z\!\Bigg)
\!\!=\!
\dot{a}\sqrt{\rho}\!\sum_{k^{\prime}=1}^{K}\!\sum_{m=1}^{M}\!u_{mk}\hat{\mathbf{g}}_{mk}^H\!\mathbf{g}_{mk^\prime}\!\sqrt{q_{k^\prime}}s_{k^\prime}
\nonumber
\\
&\!+\!\!&\dot{a}\!\!\sum_{m=1}^{M}\!\!u_{mk}\hat{\mathbf{g}}_{mk}^H\mathbf{n}_m\!+\!\sum_{m=1}^{M}\!u_{mk}e_{mk}^z
\!\!\!=\!\!
\dot{a}\!\underbrace{\sqrt{\rho} \mathbb{E}\left\{\!\sum_{m=1}^M\!u_{mk}\hat{\mathbf{g}}_{mk}^H{\mathbf{g}}_{mk}\sqrt{q_k}\right\}}_{\text{DS}_k}\!s_k
\nonumber\\
&\!+&
\!\dot{a}\!\underbrace{\sqrt{\rho}\!\left(\!\sum_{m=1}^M\!u_{mk}\hat{\mathbf{g}}_{mk}^H\!{\mathbf{g}}_{mk}\sqrt{q_k}\!-\!\mathbb{E}\!\left\{\!\sum_{m=1}^M\! u_{mk}\hat{\mathbf{g}}_{mk}^H{\mathbf{g}}_{mk}\sqrt{q_k}\!\right\}\right)}_{{\text{BU}_k}}\!s_k\!+ \!\sum_{k^{\prime}\neq k}^{K}\!\textcolor{black}{\dot{a}}
\nonumber
\\
 &&\!\!\!\underbrace{\sqrt{\rho}\!\sum_{m=1}^M\!\!u_{mk}\hat{\mathbf{g}}_{mk}^H{\mathbf{g}}_{mk^\prime}\sqrt{q_{k^\prime}}}_{{\text{IUI}_{kk^\prime}}}s_{k^{\prime}}\!\!+\!\textcolor{black}{\dot{a}}\underbrace{\!\sum_{m=1}^{M}\!\!u_{mk}\hat{\mathbf{g}}_{mk}^H\mathbf{n}_m}_{\text{TN}_k} 
 \!+\!
  \underbrace{\!\sum_{m=1}^{M}\!\!u_{mk}{e}_{mk}^z}_{\text{TQE}_k},
\label{rkcase3}
\end{IEEEeqnarray}
where $\text{TQE}_k$ refers to the total quantization error (TQE) at the $k$th user. Note that in cell-free Massive MIMO with $M \to \infty$, due to the channel hardening property, detection using only the channel statistics is nearly optimal. This is shown in \cite{marzetta_free16} (see Fig. 2 of reference \cite{marzetta_free16} and its discussion). Moreover, in \cite{marzetta_free16} the authors show that in cell-free Massive MIMO with $M \to \infty$, the received signal includes only the desired signal plus interference from the pilot sequence non-orthogonality. Finally, using the analysis in \cite{marzetta_free16}, the corresponding SINR of the received signal in (\ref{rkcase3}) can be defined by considering the worst-case of the uncorrelated Gaussian noise is given by (\ref{sinrdef11}) (defined at the top of the next page). Based on the SINR definition in (\ref{sinrdef11}), the achievable uplink rate of the \textit{k}th user is given in the following theorem.
\begin{small}
\begin{figure*}[t!]
\begin{IEEEeqnarray}{rCl} \label{sinrdef11}
\text{SINR}_k^{\text{Case 2}}= \dfrac{\left|\text{DS}_k\right|^2}{\mathbb{E}\left\{\left|\text{BU}_k\right|^2\right\}
+
\sum\limits_{k^\prime\ne k}^K\mathbb{E}\left\{\left|\text{IUI}_{kk^\prime}\right|^2\right\}
+
\mathbb{E}\left\{\left|\text{TN}_k\right|^2\right\}
+
\dfrac{1}{\textcolor{black}{\dot{a}}^2}\mathbb{E}\left\{\left|\text{TQE}_k\!\right|^2\right\}}.
\end{IEEEeqnarray}
\vspace{-.015cm}
\begin{small}
\vspace{-.04in}
	\begin{IEEEeqnarray}{rCl}\label{sinrcpustat}
\!\text{SINR}_k^{\text{Case 2}}
\!\approx\!
		\dfrac{N^2q_k\left(\sum_{m=1}^{M}u_{mk}\gamma_{mk}\right)^2}
		{
		\!N^2\!\sum\limits_{k^\prime\ne k}^K\!q_{k^\prime}\!\left(\!\sum\limits_{m=1}^{M}\!u_{mk}\gamma_{mk}\!\dfrac{\beta_{mk^\prime}}{\beta_{mk}}\!\right)^2\left|\pmb{\phi}_k^H\pmb{\phi}_{k^\prime}\right|^2
		\!+\!
		N\!\sum\limits_{m=1}^{M}u_{mk}\left(
				\!\dfrac{\sigma_{\dot{e}}^2\left(2\beta_{mk}-\gamma_{mk}\right)\!}{\textcolor{black}{\dot{a}}^2}\!+\!\gamma_{mk}\!\right)\!\sum\limits_{k^\prime=1}^{K}\!q_{k^\prime}\beta_{mk^\prime}\!+\dfrac{N}{\rho}\left(\!\dfrac{\sigma_{\dot{e}}^2}{\textcolor{black}{\dot{a}}^2}\!+1\right)\sum\limits_{m=1}^{M}u_{mk}\gamma_{mk}}.
	\end{IEEEeqnarray}
	\end{small}
		\vspace{-.023cm}
				\hrulefill
\end{figure*}
\end{small}
\begin{theorem}
	\label{theorem_up_quan_case3}
	Having the quantized weighted signal at the CPU and employing MRC detection at the CPU, the achievable uplink rate of the \textit{k}th user in the cell-free Massive MIMO system is $R = \log_2(1+\text{SINR}^{\text{Case 2}})$, where $\text{SINR}^{\text{Case 2}}$ is given by (\ref{sinrcpustat}) (defined at the top of this page).
\end{theorem}
{\textit{Proof:}}  Please refer to Appendix C. ~~
~~~~~~~~~~~~~~~~~~~~~~~~~~~~~~$\blacksquare$
\subsection{Required Fronthaul Capacity}
Let $\tau_f$ be the length of the uplink payload data transmission for each coherence interval, i.e., $
\tau_f = \tau_c - \tau_p,
$
where $\tau_c$ denotes the number of samples for each coherence interval and $\tau_p$ represents the length of pilot sequence. Defining the number of quantization bits as $\alpha_{m,i}$, for $i = 1,2$, corresponding to Cases 1 and 2, and $m$ refers to the $m$th AP. For Case 1, the required number of bits for each AP during each coherence interval is $2\alpha_{m,1}\times(NK+N\tau_f)$ whereas Case 2 requires $2\alpha_{m,2}\times(K\tau_f)$ bits for each AP during each coherence interval. Hence, the total fronthaul capacity required between the $m$th AP and the CPU for all schemes is defined as
\vspace{-.01in}
\begin{equation}
C_m\!=\!\left\{
\begin{array}{rl} 
\dfrac{2\left(NK+N\tau_f\right)\alpha_{m,1}}{T_c}, &\text{Case 1},\\\
\dfrac{2\left(K\tau_f\right)\alpha_{m,2}}{T_c},~~~~ &\text{Case 2},
\end{array} \right.
\label{fr}
\end{equation}
where $T_c$ (in sec.) refers to coherence time.\footnote{Exploiting the constraint $C_m \le C_{\text{fh}}$, the largest number of quantization level is equal to $\alpha_{m,1}=\left\lfloor\frac{T_c C_{\text{fh}}}{2 \left(NK+N\tau_f\right)} \right\rfloor$ and $\alpha_{m,2}=\left\lfloor\frac{T_c C_{\text{fh}}}{2 K \tau_f} \right\rfloor$.}
In the following, we present a comparison between two cases of uplink transmission. To make a fair comparison between Case 1 and Case 2, we use the same total number of fronthaul bits for both cases, that is
$
2(NK+N\tau_f) \alpha_{m,1} = 2(K\tau_f) \alpha_{m,2}.
$\footnote{Future work is needed to investigate the performance analysis for different numbers of quantization bits as well as the numbers of APs. This will be presented in \cite{our_asilomar18}.} 
\section{Proposed Max-Min Rate Scheme}
In this section, we formulate the max-min rate problem for Case 2 of uplink transmission in cell-free Massive MIMO system, where the minimum uplink rates of all users is maximized while satisfying the transmit power constraint at each user and the fronthaul capacity constraint.
Note that the same approach can be used to investigate the max-min rate problem for Case 1.
The achievable user SINR for the system model considered in the previous section \textcolor{black}{is obtained} by following a similar approach to that in \cite{marzetta_free16}. Note that the main difference between the proposed approach and the scheme in \cite{marzetta_free16} is the new set of receiver coefficients which are introduced at the CPU to improve the achievable user rates. The benefits of the proposed approach in terms of the achieved user uplink rate is demonstrated through numerical simulation results in Section V. In deriving the achievable rates of each user, it is assumed that the CPU exploits only the knowledge of channel statistics between the users and APs to detect data from the received signal in (\ref{rkcase3}). Using the SINR given in (\ref{sinrcpustat}), the achievable rate is obtained
$R_k^{\text{UP}} = \log_2(1 + \text{SINR}_k^{\text{Case 2}})$. Defining $\mathbf{u}_k=\left[u_{1k}, u_{2k}, \cdots, u_{Mk}\right]^T$, $\bgama_k=\left[\gamma_{1k}, \gamma_{2k}, \cdots, \gamma_{Mk}\right]^T$, $\bup_{kk^\prime}\!=\text{diag}\Big[\!\beta_{1k^\prime}\left(\frac{\sigma_{\dot{e}}^2\left(2\beta_{1k}\!-\!\gamma_{1k}\!\right)\!}{\textcolor{black}{\dot{a}}^2}\!+\!\gamma_{1k}\right),\!\cdots$,
$\beta_{Mk^\prime}\left(\frac{\sigma_{\dot{e}}^2\left(2\beta_{Mk}-\gamma_{Mk}\!\right)\!}{\textcolor{black}{\dot{a}}^2}\!+\!\gamma_{Mk}\right)\!\Big] \nonumber$, $\blam_{k k^\prime}=\left[\dfrac{\gamma_{1k}\beta_{1k^\prime}}{\beta_{1k}}, \dfrac{\gamma_{2k}\beta_{2k^\prime}}{\beta_{2k}}, \cdots, \dfrac{\gamma_{Mk}\beta_{Mk^\prime}}{\beta_{Mk}}\right]^T$ and
\\$\mathbf{R}_{k} = \text{diag}\left[\left(\dfrac{\sigma_{\dot{e}}^2}{\textcolor{black}{\dot{a}}^2}+1\right)\gamma_{1k}, \cdots, \left(\dfrac{\sigma_{\dot{e}}^2}{\textcolor{black}{\dot{a}}^2}+1\right)\gamma_{Mk}\right]$, the achievable uplink rate of the \textit{k}th user is given by
\begin{figure*}[b!]
		\vspace{-.05cm}
				\hrulefill
				\vspace{-.02in}
\begin{IEEEeqnarray}{rCl}
R_k^{\text{UP}} \approx \log_2\left( 1+\dfrac{\mathbf{u}_k^H\left(N^2q_k\bgama_k\bgama_k^H\right)\mathbf{u}_k}{\mathbf{u}_k^H\left(N^2\sum_{k^\prime\ne k}^Kq_{k^\prime}|\pmb{\phi}_k^H\pmb{\phi}_{k^\prime}|^2\blam_{k k^\prime}\blam_{k k^\prime}^H+N\sum_{k^\prime=1}^{K}q_{k^\prime}\bup_{kk^\prime}+\dfrac{N}{\rho}\mathbf{R}_{k}\right )\mathbf{u}_k}\right).~~~~
\label{sinr1}
\end{IEEEeqnarray}
\end{figure*}
\label{secop}
Next, the max-min rate problem can be formulated as follows:
\vspace{-.04in}
\begin{subequations}
\label{p1} 
\begin{align}P_1:~~
\label{p1_1}&\max_{q_k, \mathbf{u}_k,\textcolor{black}{\alpha_2}}~ \min_{k=1,\cdots,K}\quad  R_k^{\text{UP}} \\
\label{p1_2}&\text{subject to ~} ||\mathbf{u}_k||=1,  \forall k,~0 \le q_k \le p_{\text{max}}^{(k)},  ~ \forall k,\\
\label{p1_3}&~~~~~~~~~~~~~~C_m \le C_{\text{fh}},  ~~ \forall m,
\end{align}
\end{subequations}
where $p_{\text{max}}^{(k)}$ and $C_{\text{fh}}$ refer to the maximum transmit power available at user $k$ and the capacity of fronthaul link between each AP and the CPU, respectively. Note that using (\ref{fr}), $C_m$ is given as
$
C_m = \frac{2(K\tau_f)\textcolor{black}{\alpha_{m,2}}}{T_c}, \forall m.
$ \textcolor{black}{Throughout the rest of the paper, the index $m$ is dropped from $\alpha_{m,i}, ~i=1,2,$ as we consider the same number of bits to quantize the signal at all APs.}
Problem $P_{1}$ is a discrete optimization with integer decision variables and it is obvious that the achievable user rates monotonically increase with the capacity of the fronthaul link between the $m$th AP and the CPU. Hence, the optimal solution is achieved when $C_m = C_{\text{fh}},\forall m$, which leads to fixed values for the number of \textcolor{black}{quantization bits}. As a result, the max-min based max-min rate problem can be re-formulated as follows:
\vspace{-.011in}
\begin{subequations}
\label{p2} 
\begin{align}P_2:~~~~~
\label{p2_1}\max_{q_k, \mathbf{u}_k}~~~~ &\min_{k=1,\cdots,K}\quad  R_k^{\text{UP}} 
\\
\label{p2_2}~~~~~~~~~~~~\text{subject to}\quad &||\mathbf{u}_k||=1, ~ \forall k,~~~\\
\label{p2_3}&0 \le q_k \le p_{\text{max}}^{(k)},  ~~ \forall k.
\end{align}
\end{subequations}
Problem $P_{2}$ is not jointly convex in terms of $\mathbf{u}_{k}$ and power allocation $q_{k},\forall k$. Therefore, it cannot be directly solved through existing convex optimization software. To tackle this non-convexity issue, we decouple Problem $P_{2}$ into two sub-problems: receiver coefficient design (i.e. $\mathbf{u}_{k}$) and the power allocation problem. The optimal solution for Problem $P_{2}$, is obtained through alternately solving these sub-problems, as explained in the following subsections.
\subsection{Receiver Filter Coefficient Design}
In this subsection, the problem of designing the receiver coefficients is considered. We solve the max-min rate problem for a given set of allocated powers at all users, $q_k,\forall k$, and fixed values for the number of quantization levels, $Q_m,\forall m$. These coefficients (i.e., $\mathbf{u}_{k}$, $\forall k$) are obtained by independently maximizing the uplink SINR of each user. Therefore, the optimal receiver filter coefficients can be determined by solving the following optimization problem:
\begin{subequations}
\label{p3} 
\begin{align}
&P_3:~ \max_{\mathbf{u}_k} \\ \nonumber
&\dfrac{N^2\mathbf{u}_k^H\left(q_k\bgama_k\bgama_k^Hz\right)\mathbf{u}_k}
{\mathbf{u}_k^H\!\left(\!N^2\sum_{k^\prime\ne k}^K\!q_{k^\prime}|\pmb{\phi}_k^H\pmb{\phi}_{k^\prime}|^2\blam_{k k^\prime}
\blam_{k k^\prime}^H
\!+\!
N\sum_{k^\prime=1}^{K}q_{k^\prime}\bup_{kk^\prime}
\!+\!
\dfrac{N\mathbf{R}_{k}}{\rho}\!\right)\!\mathbf{u}_k}&\\
\label{p3_1}&\text{~~~~~~~~~subject to ~~~~~~}||\mathbf{u}_k||=1,~~~\forall k.&
\end{align}
\end{subequations}
Problem $P_{3}$ is a generalized eigenvalue problem \cite{bookematrix,our_lett_18,ourjournal1}, where the optimal solutions can be obtained by determining the generalized eigenvector of the matrix pair $\mathbf{A}_{k} = N^2q_k\bgama_k\bgama_k^H$ and $\mathbf{B}_{k}\!=N^2\sum_{k^\prime\ne k}^K\!q_{k^\prime}\!|\pmb{\phi}_k^H\pmb{\phi}_{k^\prime}|^2\!\blam_{k k^\prime}\!\blam_{k k^\prime}^H\!+\!N\sum_{k^\prime=1}^{K}q_{k^\prime}\bup_{kk^\prime}\!+\!\frac{N}{\rho}\!\mathbf{R}_{k}$ corresponding to the maximum generalized eigenvalue.
\vspace{-.08in}
\subsection{Power Allocation} 
In this subsection, we solve the power allocation problem for a given set of fixed receiver filter coefficients, $\mathbf{u}_{k}$, $\forall k$, and fixed values of quantization levels, $Q_m,\forall m$. The optimal transmit power can be determined by solving the following max-min problem:
\begin{subequations}
\label{p4} 
\begin{align}P_4:
\label{p4_1}\max_{q_k}~~ \min_{k=1,\cdots,K}\quad & \text{SINR}_k^{\text{UP}} \\
\label{p4_2}\text{subject to ~}\quad &0 \le q_k \le p_{\max}^{(k)}.~~~~~~
\end{align}
\end{subequations}
Without loss of generality, Problem $P_4$ can be rewritten by introducing a new slack variable as
\begin{subequations}
\label{p5} 
\begin{align}P_5:
\label{p5_1}\max_{t,q_k}\quad & t \quad &\\
\label{p5_2}\text{subject to }\quad &0\le q_k \le p_{\max}^{(k)}, ~~ \forall k, \text{SINR}_{k}^{\text{UP}} \ge t,~~ \forall k.
\end{align}
\end{subequations}
\begin{proposition}\label{prop_p5_standard}
Problem $P_{5}$ can be formulated into a standard GP. 
\end{proposition}
{\textit{Proof:}} Please refer to Appendix D.~~~~~~~~~~~~~~~~~
~~~~~~~~~~~~$\blacksquare$
\begin{algorithm}[t]
\caption{Proposed algorithm to solve Problem $P_2$}

\hrulefill

\textbf{1.} Initialize $\mathbf{q}^{(0)}=[q_1^{(0)},q_2^{(0)},\cdots,q_K^{(0)}]$, $i=1$

\textbf{2.} Repeat steps 3-5 until $\dfrac{\text{SINR}_{k}^{\text{UP},\textcolor{black}{(i)}}-\text{SINR}_{k}^{\text{UP},\textcolor{black}{(i-1)}}}{\text{SINR}_{k}^{\text{UP},\textcolor{black}{(i-1)}}}\le \epsilon, \forall k$

\textbf{3.} Determine the optimal receiver coefficients $\mathbf{U}^{(i)}=[\mathbf{u}^{(i)}_1,\mathbf{u}^{(i)}_2,\cdots,\mathbf{u}^{(i)}_K]$ through solving the generalized eigenvalue Problem $P_3$ in (\ref{p3}) for a given $\mathbf{q}^{(i-1)}$, 

\textbf{4.} Compute $\mathbf{q}^{(i)}$ through solving Problem $P_5$ in (\ref{p5}) for a given $\mathbf{U}^{(i)}$

\textbf{5.} $i=i+1$

\hrulefill
\label{al1}
\end{algorithm}
Therefore, Problem $P_5$ is efficiently solved through existing convex optimization software. Based on these two sub-problems, an iterative algorithm has been developed by alternately solving both sub-problems at each iteration. The proposed algorithm is summarized in Algorithm \ref{al1}. Note that $\epsilon$ in Step 2 of Algorithm \ref{al1} refers to a small predetermined value.
\section{Convergence}
In this section, we present the convergence of the proposed Algorithm \ref{al1}. We propose to alternatively solve two sub-problems to find the solution of the original Problem $P_2$, where at each iteration, one of the design parameters is determined by solving the corresponding sub-problem while other design variable is fixed. We showed that each sub-problem provides an optimal solution for the other given design variable. Let us assume at the $\textcolor{black}{i-1}$th iteration, that the receiver filter coefficients $\mathbf{u}_{k}^{\textcolor{black}{(i-1)}},\forall k$ are obtained for a given power allocation $\mathbf{q}^{\textcolor{black}{(i-1})}$ and similarly, the power allocation $\mathbf{q}^{\textcolor{black}{(i)}}$ is determined for a fixed set of receiver filter coefficients $\mathbf{u}_{k}^{\textcolor{black}{(i-1)}},\forall k$. 
Note that, the optimal power allocation $\mathbf{q}^{\textcolor{black}{(i)}}$
 determined
for a given $\mathbf{u}_k^{\textcolor{black}{(i-1)}}$ achieves an uplink rate greater
than or equal to that of the previous iteration. In addition, the power allocation $\mathbf{q}^{\textcolor{black}{(i-1)}}$ is a feasible solution to find $\mathbf{q}^{\textcolor{black}{(i)}}$ as the receiver filter coefficients $\mathbf{u}_{k}^{\textcolor{black}{(i)}},~ \forall k$ are determined for a given  $\mathbf{q}^{\textcolor{black}{(i-1)}}$. Note that the uplink rate of the system monotonically increases with the power. As a result, the achievable uplink rate of the system monotonically increases at each iteration.
Note that the achievable uplink max-min rate is bounded from above for a given set of per-user power constraints and fronthaul link capacity constraint. Hence the proposed algorithm converges to a specific solution. 
Note that to the best of our knowledge and referring to \cite{eehienfree} this is a common way to show the convergence.
In the next section, we prove the optimality of the proposed Algorithm \ref{al1} through the principle of uplink-downlink duality.
\vspace{-.09cm}
\section{Optimality of the Proposed Max-Min Rate algorithm}
In this section, we present a method to prove the optimality of the proposed Algorithm \ref{al1}. The proof is based
on two main observations: we first demonstrate that the original max-min Problem $P_2$ with per-user power constraint is equivalent to an uplink problem with an equivalent total power constraint. We next prove that the same SINRs can be achieved in both the uplink and the virtual downlink with an equivalent total power constraint, which enables us to establish an uplink-downlink duality. Finally, we show that the virtual downlink problem is quasi-convex and can be optimally solved through a bisection search \cite{bookboyd}. Note that the uplink Problem $P_1$ and the equivalent virtual downlink problem achieve the same SINRs and the solution of the virtual downlink problem is optimal. As a result, the optimality of the proposed Algorithm \ref{al1} is guaranteed. The details of the proof are provided in the following subsections.
\vspace{-.09cm}
\subsection{Equivalent Max-Min Uplink Problem}
We aim to show the equivalence of Problem $P_2$ with a per-user power constraint and the uplink max-min rate problem with a total power constraint. Note that in the total power constraint, the maximum available transmit power is defined as the sum of all users' transmit power from the solution of Problem $P_2$, which is formulated as:\begin{subequations}
\label{p6} 
\begin{align}P_6:~~~~
\label{p6_1}\max_{q_k, \mathbf{u}_k}~~~ &\min_{k=1,\cdots,K}\quad  R_k^{\text{UP}}~~~~\\
\label{p6_2}\text{subject to }\quad &||\mathbf{u}_k||=1, ~~ \forall k,~~~~\\
\label{totalp}&\sum_{k=1}^{K}q_k \le P_{\text{\text{tot}}}^{c}.~~~~
\end{align}
\end{subequations}
Problem $P_6$ is not convex in terms of receiver filter coefficients $\mathbf{u}_{k}$ and power allocation $q_k, \forall k$. To deal with this non-convexity, similar to the proposed method to solve problem $P_2$, we propose to modify Algorithm \ref{al1} to incorporate the total power constraint in Problem $P_6$. Hence, we decompose Problem $P_6$ into receiver filter coefficient design and power allocation sub-problems. The same generalized eigenvalue problem in Problem $P_3$ is solved to determine the receiver filter coefficients whereas the GP formulation in $P_5$ is modified to incorporate the total power constraint (\ref{totalp}).
Note that, the total power constraint is a convex constraint (posynomial function in terms of power allocation) and GP with the equivalent total power constraint can be used to find the optimum solution.
\begin{lemma}
\label{lemma_p2_p6}
The original Problem $P_2$ (with per-user power constraint) and the equivalent Problem $P_6$ (with the equivalent total power constraint) have the same optimal solution.
\end{lemma}
\vspace{-.04cm}
\textit{Proof: Please refer to Appendix E.~~~~~~~~~~~~~~~~~~
~~~~~~~~~~~~~~~~$\blacksquare$}
\vspace{-.075cm}
\subsection{Uplink-Downlink Duality for Cell-free Massive MIMO}\label{section_duality}
This subsection demonstrates an uplink-downlink duality for cell-free Massive MIMO systems. In particular, it is shown that the same SINRs (or rate regions) can be realized for all users in the uplink and the virtual downlink with the equivalent total power constraints \cite{tse, boche_sol5}, respectively. In other words, based on the principle of uplink-downlink
duality, the same set of filter coefficients can be utilized in the uplink and the downlink to achieve the same SINRs for all users with different user power allocations. The following theorem defines the achievable virtual downlink rate for cell-free Massive MIMO systems: 
\vspace{-.031in}
\begin{theorem}
\label{theorem_quan_updl}
By employing conjugate beamforming at the APs, the achievable virtual downlink rate of the $k$th user in the cell-free Massive MIMO system with $K$ randomly distributed single-antenna users, $M$ APs where each AP is equipped with $N$ antennas and limited-capacity fronthaul links is given by (\ref{sinrdlquan}) (defined at the top of this page).
\end{theorem}
\begin{figure*}[!t]
\begin{IEEEeqnarray}{rCl}
\vspace{-.0818in}
\begin{split}
\text{SINR}_k^{\text{DL}}(\mathbf{U},\mathbf{p}) = 
\dfrac{\mathbf{u}_k^H\left(N^2p_k\bgama_k\bgama_k^H\right)\mathbf{u}_k}{N^2\sum_{k^\prime\ne k}^K\mathbf{u}_{k^\prime}^Hp_{k^\prime}|\pmb{\phi}_{k^\prime}^H\pmb{\phi}_{k}|^2\bdel_{k^\prime k}\bdel_{k^\prime k}^H\mathbf{u}_{k^\prime}+N\sum_{k^\prime=1}^{K}\mathbf{u}_{k^\prime}^Hp_{k^\prime}\mathbf{F}_{k^\prime k}\mathbf{u}_{k^\prime}+\frac{N}{\rho}}.
\label{sinrdlquan}
\end{split}
\end{IEEEeqnarray}
\vspace{-.062cm}
\begin{IEEEeqnarray}{rCl}
\begin{split}
\text{SINR}_k^{\text{UP}}(\mathbf{U},\mathbf{q}) = \dfrac{\mathbf{u}_k^H\left(N^2q_k\bgama_k\bgama_k^H\right)\mathbf{u}_k}{\mathbf{u}_k^H\left(N^2\sum_{k^\prime\ne k}^Kq_{k^\prime}|\pmb{\phi}_k^H\pmb{\phi}_{k^\prime}|^2\blam_{k k^\prime}\blam_{k k^\prime}^H+N\sum_{k^\prime=1}^{K}q_{k^\prime}\bup_{kk^\prime}+\dfrac{N}{\rho}\mathbf{R}_{k}\right )\mathbf{u}_k}.
\label{sinrupquan}
\end{split}
\vspace{-.051in}
\end{IEEEeqnarray}
\vspace{-.061in}
\hrulefill
\vspace{-.051cm}
\end{figure*}
{\textit{Proof:}} This can be derived by following the same approach as for uplink transmission in Theorem \textcolor{black}{2}.~~~~
~~~~~~~~~~~~~~~~~~~~~~~~~$\blacksquare$
\\\
Note that in (\ref{sinrdlquan}), $p_k, ~ \forall k$ denotes the downlink power allocation for the $k$th user and the following equalities hold: $\bgama_k=\left[\gamma_{1k}, \gamma_{2k}, \cdots, \gamma_{Mk}\right]^T,$
$\mathbf{F}_{k^\prime k}\!=\text{diag}\Big[\!\beta_{1k}\left(\frac{\textcolor{black}{\dot{a}}^2\left(2\beta_{1k^\prime}\!-\!\gamma_{1k^\prime}\!\right)\!}{\sigma_{\dot{e}}^2}\!+\!\gamma_{1k^\prime}\right),\!\cdots\!,$
$\beta_{Mk}\left(\frac{\textcolor{black}{\dot{a}}^2\left(2\beta_{Mk^\prime}-\gamma_{Mk^\prime}\!\right)\!}{\sigma_{\dot{e}}^2}\!+\!\gamma_{Mk^\prime}\right)\!\Big] \nonumber$ and
$\bdel_{k^\prime k}=\left[\dfrac{\gamma_{1k^\prime}\beta_{1k}}{\beta_{1k^\prime}}, \dfrac{\gamma_{2k^\prime}\beta_{2k}}{\beta_{2k^\prime}}, \cdots, \dfrac{\gamma_{Mk^\prime}\beta_{Mk}}{\beta_{Mk^\prime}}\right]^T$.
The following Theorem provides the required condition to establish the uplink-downlink duality for cell-free Massive MIMO systems with limited-capacity fronthaul links:
\vspace{-.03in}
\begin{theorem}
\label{theorem_quan_updl_power}
By employing MRC detection in the uplink and conjugate beamforming in the virtual downlink, to realize the same SINR tuples in both the uplink and the virtual downlink of a cell-free Massive MIMO system, with the same fronthaul loads, the same filter coefficients and different transmit power allocations, the following condition should be satisfied:
\begin{IEEEeqnarray}{rCl}
\label{dualp}
N\sum_{m=1}^{M}\sum_{k=1}^{K}\left(\frac{\sigma_{\dot{e}}^2}{\textcolor{black}{\dot{a}}^2}+1\right)\gamma_{mk}\left| w_{mk}\right|^2= \sum_{k=1}^{K} q_k^*=P_{\text{\text{tot}}}^{c} ~,
\end{IEEEeqnarray}
\textcolor{black}{where $q_k^*,~ \forall k$ refer to the optimal solution of Algorithm 1}, and $w_{mk}$ denotes the $(m,k)$-th entry of matrix $\mathbf{W}$  which is defined as follows:
\begin{IEEEeqnarray}{rCl}
\label{wei}
\mathbf{W}=[\sqrt{p_1}\mathbf{u}_1,\sqrt{p_2}\mathbf{u}_2,\cdots,\sqrt{p_K}\mathbf{u}_K].
\end{IEEEeqnarray}
\end{theorem}
{\textit{Proof:}} Please refer to Appendix F.~~~
~~~~~~~~~~~~~
~~~~~~~~~~~~~~~~$\blacksquare$
\subsection{Equivalent Max-Min Downlink Problem}
In this subsection, we provide an optimal solution for the max-min rate downlink problem with the equivalent total power constraint. This problem can be written as follows:
\begin{subequations}
\label{p7} 
\begin{align}P_7:~~~~
\label{p7_1}\max_{p_k, \mathbf{u}_k} &\min_{k=1,\cdots,K}\quad  R_k^{\text{DL}}~~~~~~~~~\\
\label{p7_2}~~~~~~~~~~~\text{subject to }\quad &||\mathbf{u}_k||=1, ~~ \forall k, \sum_{k=1}^{K}p_k \le P_{\text{\text{tot}}}^{c},~
\end{align}
\end{subequations}
where $R_k^{\text{DL}}=\log_2(1+\text{SINR}_k^{\text{DL}})$, and $\text{SINR}_k^{\text{DL}}$ is defined in (\ref{sinrdlquan}). 
This problem is difficult to jointly solve in terms of transmit filter coefficients $\mathbf{u}_k$'s and power allocations $p_k$'s. However, it can be represented by introducing a new variable $\mathbf{W}$ to decouple the variables $\mathbf{U}$ and $\mathbf{q}$ as follows:
\begin{subequations}
\label{p8} 
\begin{align}P_8:
\label{p8_1}\max_{\mathbf{W}}& \min_{k=1,\cdots,K}\quad R_k^{\text{DL}}~~~~\\
\label{p8_2}\text{subject to}\quad &N\sum_{m=1}^{M}\sum_{k=1}^{K}
\left(\frac{\sigma_{\dot{e}}^2}{\textcolor{black}{\dot{a}}^2}
+
1\right)\gamma_{mk}| w_{mk}|^2\le P_{\text{\text{tot}}}^{c}.
\end{align}
\end{subequations}
It is easy to show that Problem $P_8$ is quasi-convex. Hence, a bisection \cite{bookboyd} approach can be used to obtain the optimal solution for the original Problem $P_8$ by sequentially solving the following power minimization problem for a given target SINR $t$ at all users:
\begin{subequations}
\label{p9} 
\begin{align}
&P_9:~  \min_{{\mathbf{W}}} ~~~~\sum_{m=1}^{M}\sum_{k=1}^{K}\gamma_{mk}| w_{mk}|^2&\\
\label{p9_1}&{\text{subject}~\text{to}~ }\\
&
\small\dfrac{\mathbf{w}_k^H\left(
N^2\bgama_k\bgama_k^H\right)\mathbf{w}_k}{N^2\sum_{k^\prime\ne k}^K\mathbf{w}_{k^\prime}^H|\pmb{\phi}_{k^\prime}^H\pmb{\phi}_{k}|^2\bdel_{k^\prime k}\bdel_{k^\prime k}^H\mathbf{w}_{k^\prime}+N\sum_{k^\prime=1}^{K}\mathbf{w}_{k^\prime}^H\mathbf{F}_{k^\prime k}\mathbf{w}_{k^\prime}+\frac{N}{\rho}}\ge t,&\nonumber\\
\label{p9_2}&~~~~~~~~~~~~N\sum_{m=1}^{M}\sum_{k=1}^{K}\left(\frac{\sigma_{\dot{e}}^2}{\textcolor{black}{\dot{a}}}+1\right)\gamma_{mk}| w_{mk}|^2 \le P_{\text{tot}}^c,&
\end{align}
\end{subequations}
where $\mathbf{w}_{k}$ represents the \textit{k}th column of the matrix $\mathbf{W}$ defined in (\ref{wei}). Problem $P_{9}$ can be reformulated by exploiting a second order cone programming (SOCP). Note that the objective function in (\ref{p9}) refers to the total transmit power. As a result, the optimal solution for Problem $P_{7}$ can be obtained by extracting the normalized transmit filter coefficients $\mathbf{u}_{k}$'s and power allocations $p_{k}$'s as 
\vspace{-.024cm}
\begin{eqnarray}
&p_k^* = ||\mathbf{w}_k^*||^2,~\forall k,~~\&~~~\mathbf{u}_k^*=\dfrac{\mathbf{w}_k^*}{||\mathbf{w}_k^*||},~\forall k,
\end{eqnarray}
where the $\mathbf{w}_{k}^{*}$'s refer to the optimal solution of Problem $P_8$. Note that constraint (\ref{p9_2}) is an equivalent total power constraint to the per-user power constraint in the original Problem $P_2$, which is a more relaxed constraint than (\ref{p2_3}). However, it is already shown in the previous sub-section that the same SINRs can be realized in both the uplink and the virtual downlink with per-user and the equivalent total power constraints.
\subsection{Prove of Optimality of Algorithm \ref{al1}}
In Lemma 2, we prove that Problems $P_2$ and $P_6$ are equivalent, and have the same solution. Next, in Proposition 1, using uplink-downlink duality, we prove that Problem $P_6$ and the virtual downlink Problem $P_7$ are equivalent. Note that the SINR achieved by solving Problem $P_7$ are optimal (the optimal solution is obtained by a bisection search approach). This confirms that the proposed algorithm to solve Problem $P_2$ is optimal.
\vspace{-.04cm}
\section{User Assignment}
Exploiting (\ref{fr}), it is obvious that the total fronthaul capacity required between the $m$th AP and the CPU increases linearly with the total number of users served by the $m$th AP. This motivates the need to pick a proper set of active users for each AP. Using (\ref{fr}), we have
\vspace{-.1in}
\begin{equation}
\alpha_2\times K_{m}\le \dfrac{C_{\text{fh}}T_c}{2\tau_f},
\label{qk}
\end{equation}
where $K_m$ denotes the size of the set of active users for the $m$th AP.
From (\ref{qk}), it can be seen that decreasing the size of the set of active users 
allows for a larger number of quantization levels. Motivated by this fact, and to exploit the capacity of fronthaul links more efficiently, we investigate all possible combinations of $\alpha_2$ and $K_{m}$. First, for a fixed value of $\alpha_2$, we find an upper bound on the size of the set of active users for each AP. In the next step, we propose for all APs that the users are sorted according to $\beta_{mk},\forall k$, and find the $K_{m}$ users which have the highest values of $\beta_{mk}$ among all users. If a user is not selected by any AP, we propose to find the AP which has the best link to this user (in Algorithm \ref{al_ap_assign}, $\pi(j) = \operatornamewithlimits{argmax}\limits_{m} \beta_{mj}$ determines best link to the $j$th user, i.e., the index of the AP which is closest to the $j$th user). Note that
to only consider the users that have links to other APs, we use $k|\mathcal{S}_k\pi_j \neq \varnothing$, where $\varnothing$ refers to empty set. Then we drop the user which has the lowest $\beta_{mk},\forall k$, among the set of active users for that AP, which has links to other APs as well. Finally, we add the user which is not selected by any AP to the set of active users for this AP. We next solve the virtual downlink problem to maximize the minimum uplink rate of the users as follows
\begin{subequations}
\label{p10} 
\begin{align}P_{10}:
\label{p10_1}\max_{\mathbf{W}}& \min_{k=1,\cdots,K}\quad R_k^{\text{DL}}\left(\vec{\gamma}_{mk}\right),~~~~\\
\label{p10_2}\text{subject to\!}\quad &N\sum_{m=1}^{M}\!\sum_{k=1}^{K}\!\left(\!\frac{\sigma_{\dot{e}}^2}{\textcolor{black}{\dot{a}}^2}\!+\!1\!\right)\!\vec{\gamma}_{mk}\!\left| w_{mk}\!\right|^2\le P_{\text{\text{tot}}}^{c},
\end{align}
\end{subequations}
where
\begin{equation}
\vec{\gamma}_{mk}=\left\{
\begin{array}{rl}
&\!\!\!\!\!\!\gamma_{mk},~~~~~~~m\in \mathcal{S}_k\\
&\!\!\!\!\!\!0, ~~~~~~~~~~\text{otherwise}
\end{array} \right.
\end{equation}
where $\mathcal{S}_k$ refers to the set of active APs for the $k$th user.
The proposed algorithm is summarized in Algorithm \ref{al_ap_assign}.
\begin{algorithm}[t]
\caption{User Assignment}
\hrulefill

\textbf{1.} Using (\ref{qk}), find the maximum possible integer value for $K_m, \forall m$
\vspace{.02cm}

\textbf{2.} Sort users according to the ascending channel gain:
$\beta_{m1} \ge \beta_{m2} \ge\cdots \ge\beta_{mK}, \forall m$

\textbf{3.} Assign $K_m$ users with the highest values of $\beta_{mk},\forall m$ to each AP, i.e., 
$\mathcal{T}_m\gets\{{k^{(1)},k^{(2)},\cdots,k^{(K_m)}}\}, \forall m$
\vspace{.02cm}

\textbf{4.} Find set of active APs for each user; $\mathcal{S}_k\gets\{{m^{(1)},m^{(2)},\cdots,m^{(M_k)}}\}, \forall k$
\vspace{.02cm}

\textbf{5.} for $j=1:K$

if ~$\text{size}~\{\mathcal{S}_j\}~=0$

$\pi(j) = \operatornamewithlimits{argmax}\limits_{m} \beta_{mj}$, $\delta(j) = \operatornamewithlimits{argmin}\limits_{k} \beta_{\pi(j)k}, \textcolor{black}{k|\mathcal{S}_k\pi_j \neq \varnothing}$, $\mathcal{T}_{\pi(j)}\gets\mathcal{T}_{\pi(j)} \textcolor{black}{\backslash} \delta(j)$, $\mathcal{T}_{\pi(j)}\gets\mathcal{T}_{\pi(j)} \cup j$ 

~~~~~~~end

~~~end

\textbf{6.} If $m\in \mathcal{S}_k$, then $\vec{\gamma}_{mk}\gets \gamma_{mk}$, otherwise $\vec{\gamma}_{mk}=0$ and solve the max-min rate problem $P_2$~~~\\
\hrulefill
\label{al_ap_assign}
\end{algorithm}
\section{Numerical Results and Discussion}\label{sec_numer}
In this section, we provide numerical simulation results to validate the performance of the proposed max-min rate scheme with different parameters. A cell-free Massive MIMO system with $M$ APs and $K$ single-antenna users is considered in a $D \times D$ simulation area, where both APs and users are uniformly distributed at random. In the following subsections, we define the simulation parameters and then present the corresponding simulation results. 
The channel coefficients between users and APs are modeled in Section II where the coefficient $\beta_{mk}$ is given by
$
\beta_{mk} = \text{PL}_{mk}  10^{\frac{\sigma_{sh} ~z_{mk}}{10}},
$
where $\text{PL}_{mk}$ is the path loss from the $k$th user to the $m$th AP and the second term$10^{\frac{\sigma_{sh}z_{mk}}{10}}$, denotes the shadow fading with standard deviation $\sigma_{sh}=8$ dB, and $z_{mk} \sim  \mathcal{N}(0,1)$ \cite{marzetta_free16}. In the simulation, an uncorrelated shadowing model is considered and a three-slope model for the path loss similar to \cite{marzetta_free16}.
\textcolor{black}{The noise power is given by
$
p_n = \text{BW} \times k_B   \times T_0 \times W,
$}
where $\text{BW}=20$ MHz denotes the bandwidth, $k_B = 1.381 \times 10^{-23}$ represents the Boltzmann constant, and $T_0 = 290$ (Kelvin) denotes the noise temperature. Moreover, $W=9$dB, and denotes the noise figure. It is assumed that that $\bar{p}_p$ and $\bar{\rho}$ denote the power of pilot sequence and the uplink data powers, respectively, where $p_p=\frac{\bar{p}_p}{p_n}$ and $\rho=\frac{\bar{\rho}}{p_n}$. In simulations, we set $\bar{p}_p=200$ mW and $\bar{\rho}=200$ mW. Similar to \cite{marzetta_free16}, we assume that the simulation area is wrapped around at the edges which can simulate an area without boundaries. Hence, the square simulation area has eight neighbours. We evaluate the average rate of the system over 300 random realizations of the locations of APs, users and shadow fading. Similar to the model in \cite{fettwis_globe11}, the fronthaul links establish communications through wireless microwave links with limited capacity. Hence, we use $C_{\text{fh}}=100$ Mbits/s \cite{fettwis_globe11}, unless otherwise it is indicated. \textcolor{black}{In this paper, the term “orthogonal pilots” refers to the case where unique orthogonal pilots are assigned to all users, while in “random pilot assignment” each user is randomly assigned a pilot sequence from a set of $\tau_p$ orthogonal sequences of length $\tau_p$ ($<K$), following the approach of \cite{marzetta_free16}.}
\begin{figure*}[t!]
\centering
\vspace{-.024in}
\subfloat[Average per-user uplink rate for cases 1 and 2, with ($N=4$, $K=20$, $\tau_p=20$, $\alpha_{1}=9$, $\alpha_{2}=2$), and ($N=20$, $K=40$, $\tau_p=40$, $\alpha_{1}=8$, $\alpha_{2}=5$) with $D=1$ km and $\tau_c=200$. Note that here $\tau_f=\tau_c-\tau_p=160$.]{\label{revision_k40_t40_n4_n20_bit8592}\includegraphics[height=61mm]{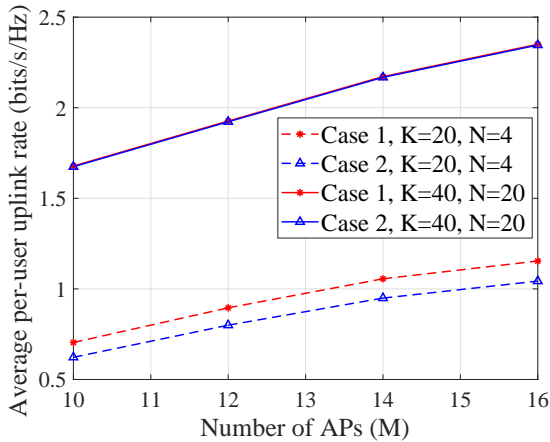}}
\hspace{1pt}
\vspace{-.09in}
\subfloat[Average per-user uplink rate for cases 1 and 2, for $M=20$, $K=20$, $\tau_p=20$, $\tau_p=10$, $D=1$ km and $\tau_c=200$ versus number of antennas per AP. Note that we consider $(\alpha_1=18, \alpha_2=5)$, $(\alpha_1=18, \alpha_2=10)$, $(\alpha_1=18, \alpha_2=15)$ for the cases of $N=5$, $N=10$, $N=15$, respectively.]{\label{revision_k20_t20_t10_m20_bit18000}\includegraphics[height=60mm]{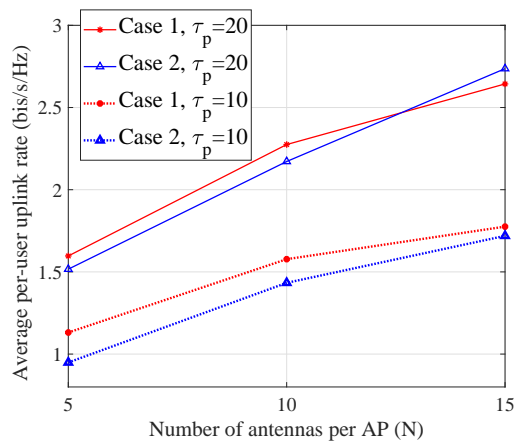}}
\label{different_cases}
\vspace{-.09in}
\caption{Performance of different cases of uplink transmission}
\vspace{-.09in}
\end{figure*}
\subsubsection{Performance of Different Cases of Uplink Transmission}Fig. \ref{revision_k40_t40_n4_n20_bit8592} presents the average per-user uplink rate, where the per-user uplink rate is obtained by solving Problem $P_4$, given by (\ref{p4}) for Cases 1 and 2.  The values of $\alpha_1=9$ and $\alpha_2=2$ correspond to a total number of 14,400 bits for each AP during each coherence time (or frame).
In addition, similar to \cite{Oppenheimsignal} we use a uniform quantizer with fixed step size.  As Fig \ref{revision_k40_t40_n4_n20_bit8592} shows the performance of Case 1 is slightly better than Case 2 for $K=20$. Next, the performance of the cell-free Massive MIMO system is evaluated for a system with $K=40$ in which each AP is equipped with $N=20$ antennas.
Fig. \ref{revision_k40_t40_n4_n20_bit8592} shows the average rate of the cell-free Massive MIMO system, where for Case 1 and Case 2, we set $\alpha_1=3$ and $\alpha_2=8$, respectively which leads to a total number of 64,000 fronthaul bits per AP per frame. Fig. \ref{revision_k40_t40_n4_n20_bit8592} shows that the performances of Case 1 and Case 2 depend on the values of $N$, $K$ and $\tau_f$. Next, we investigate the effect of number of antennas per AP and $\tau_f$ for $K=20$. Fig. \ref{revision_k20_t20_t10_m20_bit18000} shows the average per-user uplink rate of cell-free Massive MIMO versus number of antennas per AP and two cases of $\tau_p=20$ ($\tau_f=180$) and $\tau_p=10$ ($\tau_f=190$). Moreover, we consider $(\alpha_1=18, \alpha_2=5)$, $(\alpha_1=18, \alpha_2=10)$, $(\alpha_1=18, \alpha_2=15)$ for the cases of $N=5$, $N=10$, $N=15$, respectively, resulting $18,000$ bits for all values of $N$. As the figure shows the difference between Case 1 and Case 2 decreases as $N$ increases. Moreover, for the case of orthogonal pilots and $N=15$, the performance of Case 2 is better than the performance of Case 1. Since in case 1, the CPU knows the quantized channel estimates, other signal processing techniques (e.g., zero-forcing processing) can be implemented to improve the system performance and can be considered in future work. 
\begin{figure}[t!]
	\center
	\includegraphics[width=83mm]{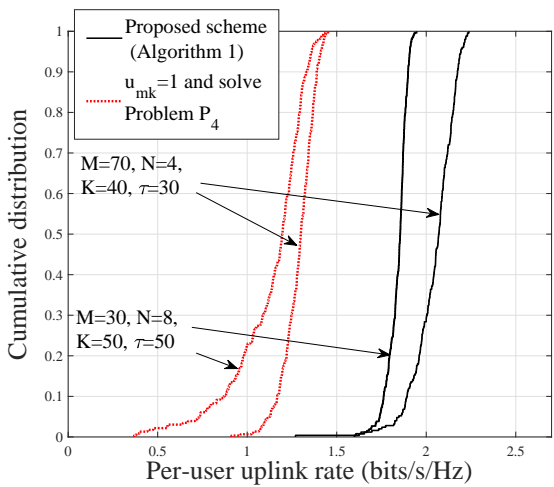}
	\vspace{-.1in}
	\caption{The cumulative distribution of the per-user uplink rate, for $\{M=70, N=4, K=40\}$, $\{M=50, N=4, K=30\}$, and $\tau_p=30$, $\alpha_1=1$ and $D = 1$ km.}
	\label{gp_bi_compare_M70_n4_k40_t30_M50_n4_k30_t10_ortho_q1_buss}
\end{figure}
\subsubsection{Performance of the Proposed User Max-Min Rate Algorithm} 
In this subsection, we evaluate the performance of the proposed uplink max-min rate scheme. To assess the performance, a cell-free Massive MIMO system is considered with 70 APs ($M=70$) where each AP is equipped with $N=4$ antennas and 40 users ($K=40$) which are randomly distributed over the simulation area of size $1 \times 1$ km meters.  Moreover, we consider the case $\{M=50, N=4, K=30\}$
Fig. \ref{gp_bi_compare_M70_n4_k40_t30_M50_n4_k30_t10_ortho_q1_buss} presents the cumulative distribution
of the achievable uplink rates for the proposed Algorithm \ref{al1} in the case similar to \cite{marzetta_free16}, without defining the coefficients $\mathbf{u}_{k}$, (i.e., ${u}_{mk}=1$ $\forall m,k$) and solving Problem $P_4$, with random pilot sequences with length $\tau_p=30$. As seen in Fig. \ref{gp_bi_compare_M70_n4_k40_t30_M50_n4_k30_t10_ortho_q1_buss}, the performance (i.e. the $10\%$-outage rate, $R_{\text{out}}$, refers to the case when $P_{\text{out}}=\text{Pr}(R_k<R_{\text{out}})=0.1$, where $\text{Pr}$ refers to the probability function) of the proposed scheme is almost three times than that of the case with ${u}_{mk}=1$ $\forall m,k$.
\begin{figure*}[t!]
\vspace{-.08in}
\centering
 \subfloat[The convergence of the proposed max-min rate approach (Algorithm \ref{al1}) for $M=70$, $N=4$, $K=40$, $\tau_p=30$, $\textcolor{black}{\alpha_2=1}$ and $D=1$ km.]{\label{conv_70404_t30_q5}\includegraphics[height=56mm]{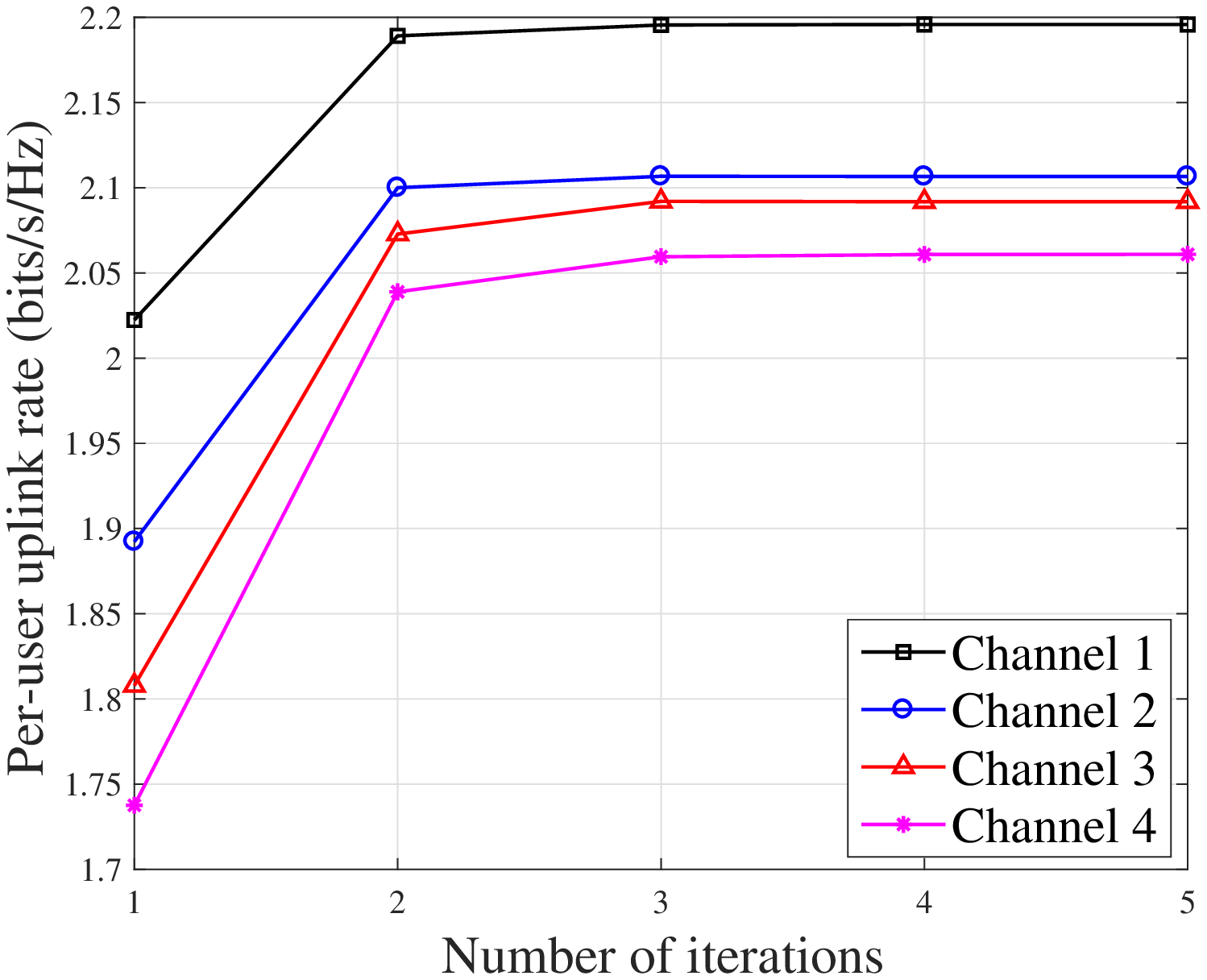}}
\hspace{1pt}
\vspace{-.11in}
\subfloat[The convergence of the proposed max-min rate approach (Algorithm \ref{al1}) for $M=30$, $N=8$, $K=50$, $\tau_p=50$, $\textcolor{black}{\alpha_2=1}$ and $D=1$ km.]{\label{conv_30808_ortho_q5}\includegraphics[height=56mm]{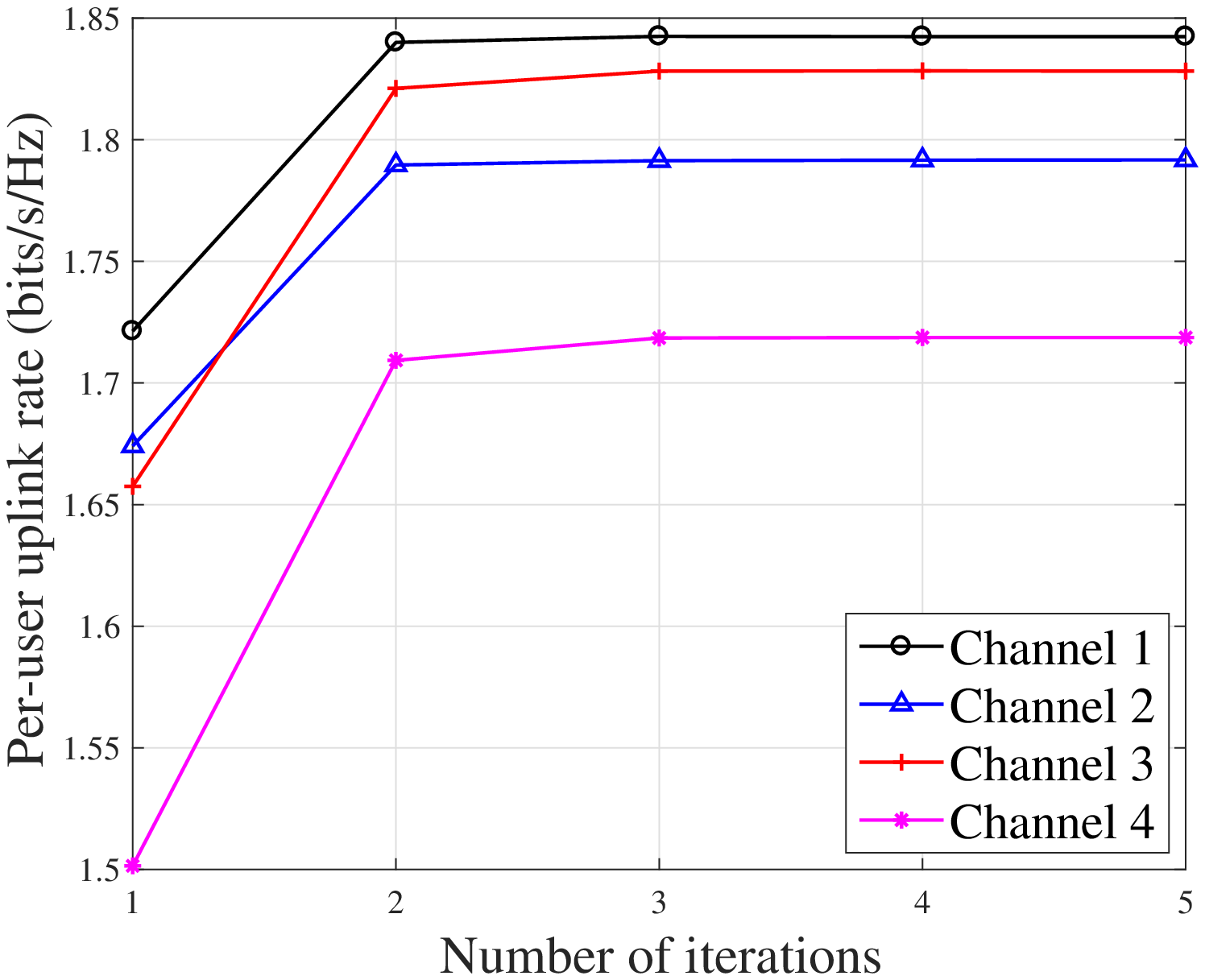}}
\label{conv}
\caption{Convergence}
\vspace{-.12in}
\end{figure*}
\subsubsection{Convergence}
Next, we provide simulation results to validate the convergence of the proposed algorithm for a set of different random realizations of the locations of APs, users and shadow fading. These results are generated over the simulation area of size $1 \times 1$ $\text{km}^2$ with random and orthogonal pilot sequences. Fig. \ref{conv_70404_t30_q5} investigates the convergence of the proposed Algorithm {\ref{al1}} with 70 APs (M = 70) and 40
users (K=40) and random
pilot sequences with length $\tau_p=30$, whereas Fig. \ref{conv_30808_ortho_q5} demonstrates the convergence of the proposed Algorithm {\ref{al1}} for the case of $M=30$ APs and $K=50$ with orthogonal pilot sequences.
The figures confirm that the proposed algorithm
converges after a few iterations, while the minimum rate of the users increases with
the iteration number.
\subsubsection{Uplink-Downlink Duality in Cell-Free Massive MIMO System}
\begin{figure}[t!]
	\center
	\includegraphics[width=78mm]{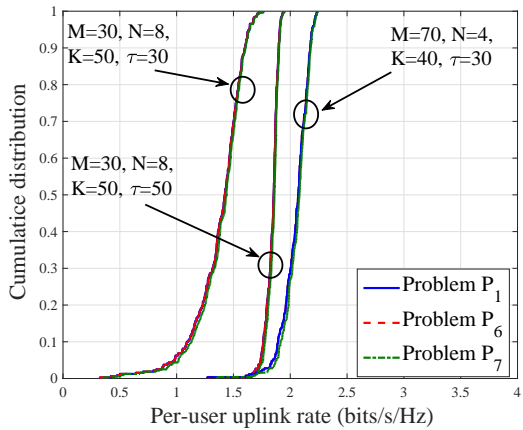}
	\vspace{-.12in}
	\caption{The cumulative distribution of the per-user uplink rate for the original problem with per-user power constraint (Problem $P_{1}$), the equivalent uplink problem with total power constraint (Problem $P_{5}$), and the equivalent downlink problem (Problem $P_{6}$), with $\alpha_1=1$ and $D = 1$ km.}
	\label{updl_70440_30850_t30ortho_q5}
	\vspace{-.12in}
\end{figure}

Here, the simulation results are provided to support the theoretical derivations of the uplink-downlink duality and the optimality of Algorithm 1. It is assumed that users are randomly distributed through the simulation area of size $1 \times 1$ km. Figs. \ref{updl_70440_30850_t30ortho_q5} compares the cumulative distribution of the achievable uplink rates between the original uplink max-min problem (Problem $P_1$), the equivalent uplink problem (Problem $P_6$) and the equivalent downlink problem (Problem $P_7$). In Fig. \ref{updl_70440_30850_t30ortho_q5}, the minimum uplink rate is obtained for a system with 30 APs ($M=30$) where each is equipped with $N=8$ antennas and has 50 users ($K=50$) for two cases of orthogonal pilot sequences and random pilot sequences with length $\tau_p=30$. Moreover, Fig. \ref{updl_70440_30850_t30ortho_q5} demonstrates the same results for 70 APs ($M=70$), $N=4$, 40 users ($K=40$), and $\tau_p=30$. The simulation results provided in Fig. \ref{updl_70440_30850_t30ortho_q5} validate our result that the problem formulations $P_1$, $P_6$ and $P_7$ are equivalent and achieve the same minimum user rate. In addition, these results support our result on the uplink-downlink duality for cell-free Massive MIMO in Section VI and the proof of optimality of Algorithm \ref{al1}.
\subsubsection{Performance of the Proposed User Assignment Algorithm \ref{al_ap_assign}}
\begin{figure*}[t!]
\vspace{-.44in}
\centering
 \subfloat[Average per-user uplink rate versus the total number of active users for each AP with  $M=120$, $N=2$, $K=50$ and $\alpha_{2}\times K_{m}=100$.]
 {\label{assign_M120_K50_tau3050_buss}\includegraphics[height=61mm]{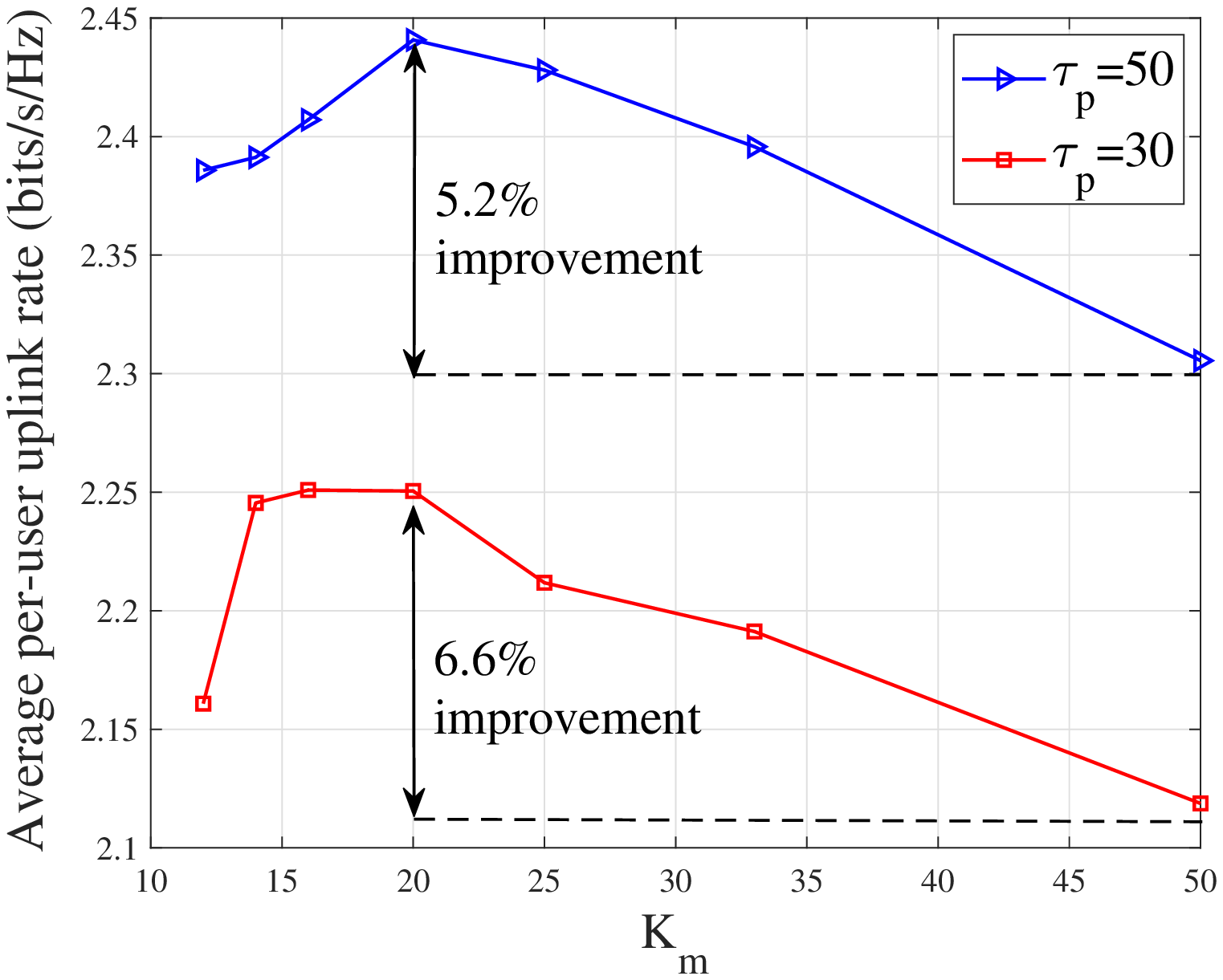}}
\hspace{1pt}
\vspace{-.08in}
\subfloat[Average per-user uplink rate versus the number of quantization bits, $\textcolor{black}{\alpha_2}$, with limited and perfect fronthaul links and $M=120$, $K=50$, $N=2$, $D=1$ km, $\tau_p=30$ and $\tau_p=50$.]{\label{backhaul_effec_m120_k50_t30_t50_assign_buss}
\includegraphics[height=61mm]{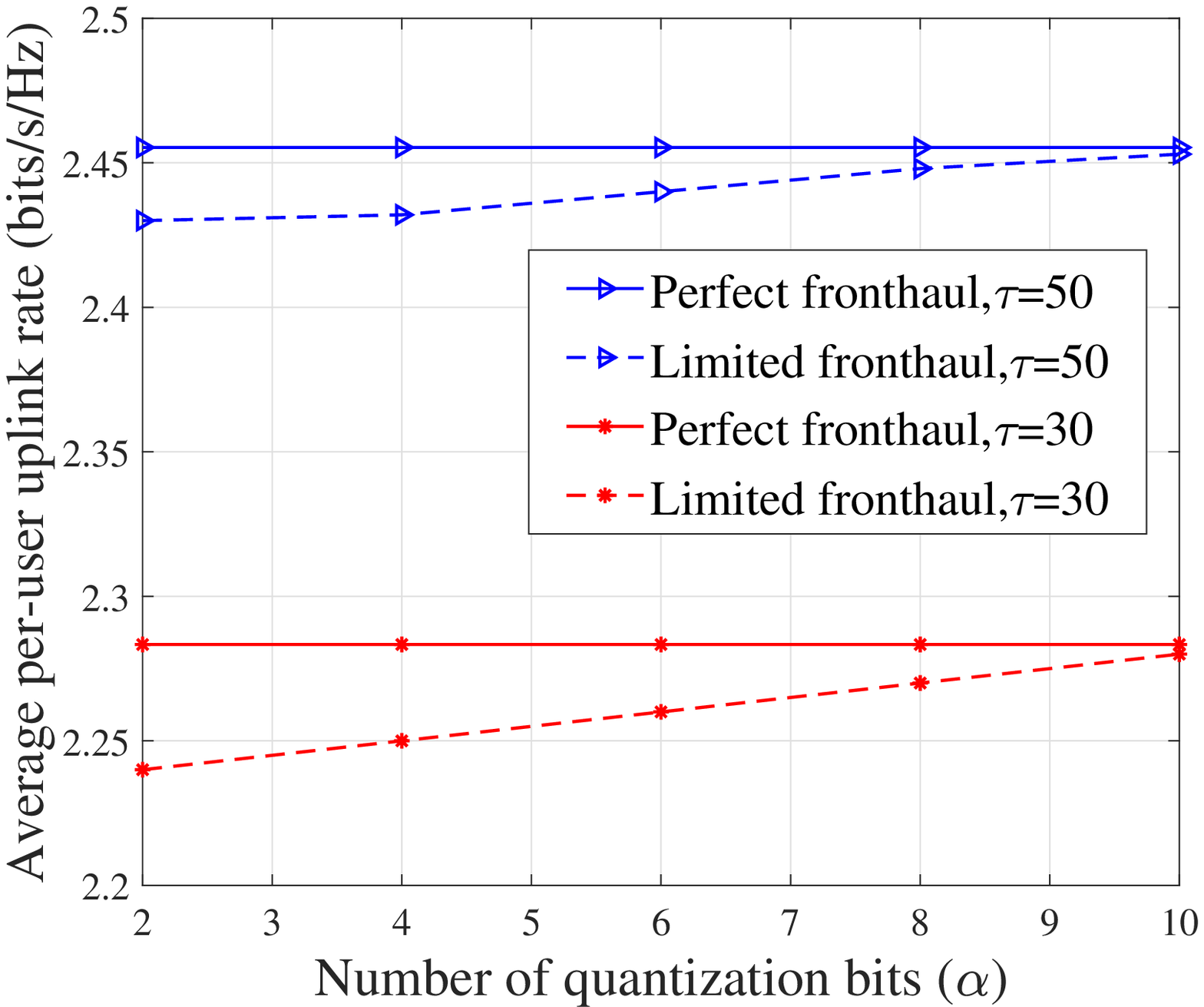}}
\caption{Average per-user uplink rate.}
\vspace{-.09cm}
\end{figure*}

This subsection investigates the performance of the proposed user assignment Algorithm \ref{al_ap_assign}. In Fig. \ref{assign_M120_K50_tau3050_buss}, the average per-user uplink rate is presented with $M=120$, $N=2$, $K=50$, orthogonal pilot sequences and random pilot assignment with $D=1$ km, versus the total number of active users per AP. Here, we used inequality (\ref{qk}) and set $\alpha_2\times K_{m}=100$ for all curves in Fig. \ref{assign_M120_K50_tau3050_buss}. The optimum value of $K_m$, ($K_m^{\text{opt}}$), depends on the system parameters and as Fig. \ref{assign_M120_K50_tau3050_buss} shows for both cases of $\tau_p=50$ and $\tau_p=30$, the optimum value is achieved by $K_m^{\text{opt}}=20$. As a result, the proposed user assignment scheme can efficiently improve the performance of cell-free Massive MIMO systems with limited fronthaul capacity. 
For instance, using  the proposed user assignment scheme for the case of $\tau_p=50$ in Fig. \ref{assign_M120_K50_tau3050_buss}, one can achieve per-user uplink rate of $2.442$ $\text{bits}/\text{s}/\text{Hz}$ by setting $K_m^{\text{opt}}=20$, instead of quantizing the signals of all $K=40$ users and achieving per-user uplink rate of $2.3$ $\text{bits}/\text{s}/\text{Hz}$, which indicates more than $5.2\%$ in the performance of cell-free Massive MIMO systems with limited fronthaul capacity.
\vspace{-.02in}
\subsubsection{Effect of the Capacity of Fronthaul Links}
What is the optimal capacity of fronthaul links in cell-free Massive MIMO systems to approach the performance of the system with perfect and error-free fronthaul links? The aim of this subsection is to answer this fundamental question. In this subsection, we evaluate the performance of the cell-free Massive MIMO system with two cases of perfect and limited fronthaul links. To assess the performance, a cell-free Massive MIMO system is considered with $M=120$, $K=50$, $N=2$, $D=1$ km, $\tau_p=30$ and $\tau_p=50$. To improve the performance of the system, we exploit the proposed user assignment algorithm.  
Fig. \ref{backhaul_effec_m120_k50_t30_t50_assign_buss} presents average per-user uplink rate with the proposed max-min rate algorithm versus number of quantization bits, $\alpha_1$ with the use of proposed user assignment algorithm. As Fig. \ref{backhaul_effec_m120_k50_t30_t50_assign_buss} shows, for both cases of random and orthogonal pilots to closely approach the performance of perfect fronthaul links, we need to set $\alpha_1\ge 8$. 
\vspace{-.4cm}
\section{Conclusions}
\vspace{-.2cm}
We have studied the uplink max-min rate problem in cell-free Massive MIMO with the realistic assumption of limited-capacity fronthaul links, and have proposed an optimal solution to maximize the minimum user rate. The original max-min problem was divided into two sub-problems which were iteratively solved by formulating them into generalized eigenvalue problem and GP. The optimality of the proposed solution has been validated through establishing an uplink-downlink duality. Numerical results have been provided to demonstrate the optimality of the proposed scheme in comparison with the existing schemes. In addition, these results confirmed that the proposed max-min rate algorithm can increase 
the median of the CDF of the minimum uplink rate of the users by more than two times, compared to existing algorithms. We finally showed that further improvement (more than three times) in minimum rate of the users can be achieved by the proposed user assignment algorithm.
\vspace{-.18cm}
\section*{Appendix A: Proof of Lemma \ref{lemmaab}} 
\vspace{-.14cm}
We exploit (\ref{ee_bussa}) and (\ref{ee_busshz}) to find $a$ and $b$ for uniform quantizer as follows:
\begin{IEEEeqnarray}{rCl}\label{ee_bussa_proof}
\vspace{-.1in}
a 
&=&
 \frac{1}{p_z}\int_{-\infty}^{\infty}zh(z)f_z(z)dz
=
\frac{1}{p_z}\left(\int_{-\infty}^{-\frac{L}{2}+1}-z\frac{L-1}{2}\Delta f_z(z)dz
\right.
\nonumber\\
	&+&\left.
\sum_{l=-\frac{L}{2}+1}^{l=-\frac{L}{2}+1}\int_{l\Delta}^{(l+1)\Delta}x\left(l+\frac{1}{2}\right)\Delta f_z(z)dz+\int_{\frac{L}{2}-1}^{\infty}z\frac{L-1}{2}\Delta f_z(z)dz\right)
\nonumber
\\
	&\stackrel{a_1}{=}&
	\frac{(L-1)\Delta}{2\sqrt{2\pi p_z}}\exp \left(-\frac{\left(\frac{L}{2}-1\right)^2}{2} \right)
	+\sum_{l=-\frac{L}{2}+1}^{\frac{L}{2}-2}
	\frac{\left(l+\frac{1}{2}\right)\Delta}{\sqrt{2\pi p_z}} 
\nonumber	
	\\
&& \!\!\!\!\!\!\!\!
\left(\exp\left(-\frac{l^2\Delta^2}{2}\right)-\exp\left(\frac{\left(l+1\right)^2\Delta^2}{2}\right)\right)+\frac{\left(L-1\right) \Delta}{2\sqrt{2\pi p_z}}
\exp\left(-\frac{\left(\frac{L}{2}-1\right)^2}{2}\right)
\nonumber 
\\
&\stackrel{a_2}{=}
\!&\!
\sum_{l=-\frac{L}{2}+1}^{\frac{L}{2}-2}\!\frac{\left(l+\frac{1}{2}\right)\Delta}{\sqrt{2\pi p_z}}\exp\left(-\frac{l^2\Delta^2}{2}\right)\!-\!\sum_{l^\prime
\!=\!-\frac{L}{2}+1}^{\frac{L}{2}-2}\frac{\left(l^\prime+\frac{1}{2}\right)\Delta}{\sqrt{2\pi p_z}}\exp\!\left(\!-\frac{l^{\prime 2}\Delta^2}{2}\right)
\nonumber
\\
&\!\!\!\!\!\!\!=\!&
\!\sum_{l=-\frac{L}{2}\!+\!1}^{\frac{L}{2}-2}\frac{\Delta}{\sqrt{2\pi p_z}}\exp\left(\!\frac{-l^2\Delta^2}{2}\!\right)
\!=\!
\sum_{l=1}^{\frac{L}{2}\!-1}\frac{2\Delta}{\sqrt{2\pi}}\exp\left(\frac{-
l^2\Delta^2}{2}\right)\!\!+\!\!\frac{\Delta}{\sqrt{2\pi p_z}}
\nonumber
\\
&\!=\!&
\Delta\sqrt{\dfrac{2}{\pi p_z}}
\left(\sum_{l=1}^{\frac{L}{2}-1}\exp\left(\frac{-l^2\Delta^2}{2}\right)+1\right),
\end{IEEEeqnarray}
\vspace{-.16in}
\begin{IEEEeqnarray}{rCl}\label{ee_bussb_proof}
&&\!\!\!\!\!b\!\!=\!\! \frac{\mathbb{E}\!\left\{h^2\!\left(z\right)\right\}}{\mathbb{E}\left\{z^2\!\right\}}\!\!=\!\!\frac{1}{p_z}\!\!\int_{-\infty}^{\infty}\!\!\!h^2(z)f_z(z)dz
\!\!=\!\!
\nonumber
\frac{2}{p_z}\!\left(\!\sum_{l=1}^{\frac{L}{2}-1}\!\int_{\left(l-1\right)\Delta}^{l\Delta}\!\left(l\!-\!\frac{1}{2}\!\right)^2\!\!\Delta^2\!f_z(z)dz
\right.
\\
&\!+\!&
\left.
\!\int_{\left(\frac{L}{2}-1\right)\Delta}^{l\Delta}\!\left(\!\frac{L-1}{2}\right)^2\!\Delta^2f_z(z)dz\!
\right)
\!
\stackrel{a_1}
{=}
\!\frac{2\Delta^2}{p_z}\!\Bigg(\!\sum_{l=1}^{\frac{L}{2}-1}\left(l\!-\!\frac{1}{2}\right)^2
\Bigg(Q\left(\frac{\left(l-1\right)\Delta}{\sqrt{p_z}}\right)
\nonumber
\\&-&
Q\left(\frac{l\Delta}{\sqrt{p_z}}\right)
\Bigg)
+
	\left(\frac{L-1}{2}\right)^2Q\left(\frac{\left(L-2\right)\Delta}{2p_z}\right)	\Bigg)
	\stackrel{a_2}
	{=}
	\frac{2\Delta^2}{p_z}\Bigg(\sum_{l^\prime=0}^{\frac{L}{2}-1}
	\left(
	l^\prime+\frac{1}{2}\right)^2
	\nonumber
	\\
	&&
	Q\left(\frac{l^\prime\Delta}{p_z}\right)
	-
	\sum_{l=1}^{\frac{L}{2}-1}\left(l-\frac{1}{2}\right)^2Q
	\left(\frac{l\Delta}{p_z}\right)
	\Bigg)
=
	\frac{2\Delta^2}{p_z}\Bigg(\left(\frac{1}{2}\right)^2Q(0)
	\nonumber
	\\
	&+&
	\!\sum_{l=1}^{\frac{L}{2}-1}\!
	\left(\!
	\left(\!l\!+\!\frac{1}{2}\right)^2\!-\!\left(l\!-\!\frac{1}{2}\right)^2
	\right)Q\left(\frac{l\Delta}{p_z}\!\right)
	\!\Bigg)
	\!=\!\frac{\Delta^2}{p_z}\left(\frac{1}{4}\!+\!4\!\sum_{l=1}^{\frac{L}{2}-1}\!lQ\left(\frac{l\Delta}{p_z}\!\right)\!\right),\!\!\!\!\!\!\!
\end{IEEEeqnarray}
where the steps $a_1$ and $a_2$ come from the property that the input of the quantizer has the Gaussian distribution, and $l^\prime=l+1$, respectively.~~~~~~~~~~~~~~~~~~~~~~~~~~~~~~~~~~~~~~~~~~~~~~~~~~~~~~~~~~~~
~~~~~~$\blacksquare$
\section*{Appendix B: Proof of Proposition \ref{prop_mutual_uncor}} 
Terms $\mathbf{e}_{m}^y$ and $\mathbf{e}_{mk}^{\hat{g}}$ have i.i.d. random variables with zero mean \cite{Oppenheimsignal}. The value of the quantization error is uncorrelated with the input of the quantizer. This can be achieved by exploiting the Bussgang decomposition \cite{Zillmann}. In this paper, we do not address the details of Bussgang decomposition and it can be considered an an interesting future direction. As a result, we have
$
	\mathbb{E}\left\{\left[\mathbf{e}_{m}^y\right]_n\right\}=0
	~\& ~\mathbb{E}\left\{\left[\mathbf{e}_{mk}^{\hat{g}}\right]_n\right\}=0
	\label{Eem}
$, 
$
	\mathbb{E}\left\{(\mathbf{e}_{m}^y)^H\mathbf{e}_{mk}^{\hat{g}}\right\}=0\label{Eemyemg}
$,
$
	\mathbb{E}\left\{\hat{\mathbf{g}}_{mk}^H\mathbf{e}_{mk}^{\hat{g}}\right\}=0\label{Egmkemkg}
$,
$
	\mathbb{E}\left\{\mathbf{y}_{m}^H\mathbf{e}_{mk}^{\hat{g}}\right\}=0\label{Eyeg}
$,
$
	\mathbb{E}\left\{\mathbf{y}_{m}^H\mathbf{e}_{m}^{y}\right\}=0\label{Eyey}
$, and
$
	\mathbb{E}\left\{\hat{\mathbf{g}}_{mk}^H\mathbf{e}_{m}^{y}\right\}=0\label{Egey}.
$
In addition, based on \cite{eehienfree}, we have
$
	\mathbf{g}_{mk} = \hat{\mathbf{g}}_{mk} + \bar{\mathbf{g}}_{mk},
	\label{gmk}
$
where $\bar{\mathbf{g}}_{mk}$ has  i.i.d. $\mathcal{CN}(0,1)$ elements. Hence, 
$
	\mathbb{E}\left\{\mathbf{g}_{mk}^H\mathbf{e}_{m}^{y}\right\}=0 ~\&~ \mathbb{E}\left\{\mathbf{g}_{mk}^H\mathbf{e}_{mk}^{\hat{g}}\right\}=0.
	\label{Ege}
$
These result in
\vspace{-.052cm}
\begin{IEEEeqnarray}{rCl}
&&\mathbb{E}\left\{\text{TQN}_{kk^\prime}\right\}=0, \mathbb{E}\left\{\text{TQN}_{k}^g\right\}=0,\\
&&\mathbb{E}\left\{\text{TQN}_{k}^y\right\}=0, \mathbb{E}\left\{\text{TQN}_{k}^{gy}\right\}=0.
	\label{Etqn}
	\vspace{-.02in}
\end{IEEEeqnarray}
Moreover, note that as the term ${\text{DS}_k}$ is a constant, we have
$	 \mathbb{E}\left\{\text{DS}_k^H\text{TQN}_k^y\right\}=
	\text{DS}_k^H\mathbb{E}\left\{\text{TQN}_k^y\right\}=0,\label{edsk1}
$
and similarly $\mathbb{E}\left\{\text{DS}_k^H\text{TQN}_k^g\right\}=0$, $\mathbb{E}\left\{\text{DS}_k^H\text{TQN}_k^{gy}\right\}=0$, and $\mathbb{E}\left\{\text{DS}_k^H\text{TQN}_{kk^\prime}\right\}=0$. In addition, we have
\vspace{-.052cm}
\begin{IEEEeqnarray}{rCl}\label{buqtn}
	\small
	&&\!\!\!\!\!\mathbb{E}\left\{\text{BU}_k^H\text{TQN}_{kk^\prime}\!\right\}
	\!=\!\mathbb{E} \Bigg\{\Bigg(\sum_{m=1}^Mu_{mk}\hat{\mathbf{g}}_{mk}^H\mathbf{g}_{mk}\sqrt{q_k}
	\nonumber
	\\
	&-&
	\underbrace{\mathbb{E}\!\left\{\sum_{m=1}^M
	u_{mk}\hat{\mathbf{g}}_{mk}^H\mathbf{g}_{mk}\sqrt{q_k}\!\right\}}_{A_1}\!\Bigg)^H
	\sum_{m=1}^{M}u_{mk}(\mathbf{e}_{mk}^{\hat{g}})^H{\mathbf{g}}_{mk^\prime}\sqrt{q_{k^\prime}} \Bigg\} 
	\nonumber
	\\
	&\!=\!&
	\mathbb{E}\!\left\{\!\left(\sum_{m=1}^{M}u_{mk}\hat{\mathbf{g}}_{mk}^H{\mathbf{g}}_{mk}\sqrt{q_k}\!\right)^H
\left(\sum_{m=1}^{M}u_{mk}(\mathbf{e}_{mk}^{\hat{g}})^H{\mathbf{g}}_{mk^\prime}\sqrt{q_{k^\prime}}\right)\!\right\}
\nonumber
\\
&-&
\mathbb{E}\left\{\!{{A}_1}^H\!
	\left(\!\sum_{m=1}^{M}\!u_{mk}(\mathbf{e}_{mk}^{\hat{g}})^H{\mathbf{g}}_{mk^\prime}\sqrt{q_{k^\prime}}\right)\!\right\}.
\end{IEEEeqnarray}
For the first term of (\ref{buqtn}), we have
\begin{IEEEeqnarray}{rCl}
&&
\mathbb{E}\left\{\left(
\sum_{m=1}^{M}u_{mk}\hat{\mathbf{g}}_{mk}^H\!{\mathbf{g}}_{mk}\sqrt{q_k}
\right)^H
\left(\sum_{m=1}^{M}u_{mk}(\mathbf{e}_{mk}^{\hat{g}})^H{\mathbf{g}}_{mk^\prime}
\sqrt{q_{k^\prime}}\right)\right\}
	\nonumber
	\\
&=&
	\sqrt{q_{k}q_{k^\prime}}\mathbb{E}\left\{\sum_{m=1}^{M}\sum_{n=1}^{M}
	u_{mk}u_{mk}
	\hat{\mathbf{g}}_{mk}^H\mathbf{g}_{mk}(\mathbf{e}_{nk}^{\hat{g}})^H\mathbf{g}_{nk^\prime}
	\right\}=0,~~~~~~
	\label{buqtn1}
	\vspace{-.02cm}
\end{IEEEeqnarray}
where the last equality is due to $	\mathbb{E}\left\{\mathbf{g}_{mk}^H\mathbf{e}_{m}^{y}\right\}=0$,
$\mathbb{E}\left\{\mathbf{g}_{mk}^H\mathbf{e}_{mk}^{\hat{g}}\right\}=0$, and $\mathbb{E}\left\{\hat{\mathbf{g}}_{mk}^H\mathbf{e}_{mk}^{\hat{g}}\right\}=0$. For the second term of (\ref{buqtn}), as $A_1$ is a constant, and using $\mathbb{E}\left\{\mathbf{g}_{mk}^H\mathbf{e}_{mk}^{\hat{g}}\right\}=0$, we have
\vspace{-.1in}
\begin{IEEEeqnarray}{rCl}
	\mathbb{E}\left\{A_1^H
	\left(\sum_{m=1}^{M}u_{mk}(\mathbf{e}_{mk}^{\hat{g}})^H{\mathbf{g}}_{mk^\prime}\sqrt{q_{k^\prime}}\right)\right\}=0.
	\label{buqtn2}
\end{IEEEeqnarray}
Finally, using (\ref{buqtn1}) and (\ref{buqtn2}), we have
$
	\mathbb{E}\left\{\text{BU}_k^H\text{TQN}_{kk^\prime}\right\}=0.
$
Using the same approach, it is easy to show that the terms $\text{DS}_k$, $\text{BU}_k$, $\text{IUI}_{kk^\prime}$, $\text{TQN}_{kk^\prime}$, $\text{TQN}_k^g$, $\text{TQN}_k^y$, and $\text{TQN}_k^{gy}$ are mutually uncorrelated, which completes the proof of Proposition \ref{prop_mutual_uncor}.~~~~~~~~~~~~~~~~~~~~~~~~~~~~~~~~~~~~~~
~~~~~~~~~~~~~~~~~~~~~~~~~~~~$\blacksquare$
\section*{Appendix C: Proof of Theorem \ref{theorem_up_quan_case3}}
The desired signal for the user $k$ is given by
\begin{IEEEeqnarray}{rCl}
\!\!\!\text{DS}_k
\!=\!
\sqrt{\rho}\mathbb{E}\left\{\sum_{m=1}^{M}\!u_{mk}\hat{\mathbf{g}}_{mk}^H\mathbf{g}_{mk}\sqrt{q_k}\right\}
\!=\!N\sqrt{p q_k}\sum_{m=1}^{M}u_{mk}\gamma_{mk}.~~
\label{dsk_vector}
\end{IEEEeqnarray}
Hence, $\left|\text{DS}_k \right| ^2 = \rho q_k \left(N\sum_{m=1}^{M}u_{mk}\gamma_{mk}\right)^2$. Moreover, the term $\mathbb{E}\{\left | \text{BU}_k\right |^2\}$ can be obtained as
\begin{IEEEeqnarray}{rCl}
&&\mathbb{E} \left\{ \left | \text{BU}_k \right | ^2\right\} =\rho\sum_{m=1}^M q_ku_{mk}^2\left( \mathbb{E}\left\{\left | \hat{\mathbf{g}}_{mk}^H{\mathbf{g}}_{mk} - \mathbb{E} \left\{ \hat{\mathbf{g}}_{mk}^H{\mathbf{g}}_{mk}\right\} \right | ^2\right\} \right)
\nonumber
\\
&=&
\rho N\sum_{m=1}^Mq_ku_{mk}^2\gamma_{mk}\beta_{mk},~
\label{ebuk}
\end{IEEEeqnarray}where the last equality comes from the analysis in \cite[Appendix A]{marzetta_free16}, and using $\gamma_{mk}=\sqrt{\tau_p p_p}\beta_{mk}c_{mk}$.
The term $\mathbb{E}\{\left | \text{IUI}_{k k^\prime}\right |^2\}$ is obtained as 
\vspace{-.1in}
\begin{IEEEeqnarray}{rCl}
&&\mathbb{E} \left\{| \text{IUI}_{k k^\prime} |^2 \right\}
=
 \rho \underbrace{q_{k^\prime} \mathbb{E}\left \{\left |\sum_{m=1}^Mc_{mk}u_{mk}\mathbf{g}_{mk^\prime}^H\tilde{\bf{w}}_{mk}  \right |^2\right\}}_{A}
\nonumber
\\
&+&
\rho \underbrace{\tau_p p_p \mathbb{E} \left \{q_{k^\prime}\left |  \sum_{m=1}^M
c_{mk}u_{mk}\left(\sum_{i=1}^{K}\mathbf{g}_{mi}\pmb{\phi}_k^H\pmb{\phi}_i\right)^H{\mathbf{g}}_{mk^\prime}\right |^2\right \} }_{B},
 \label{eiui}
\end{IEEEeqnarray}where the third equality in (\ref{eiui}) is due to the fact that for two independent random variables $X$ and $Y$ and $\mathbb{E}\{X\}=0$, we have $\mathbb{E}\{\left | X+Y \right |^2\}=\mathbb{E}\{\left | X \right |^2\}+\mathbb{E}\{\left | Y \right |^2\}$ \cite{marzetta_free16}.
Since $\tilde{\bf{w}}_{mk}=\pmb{\phi}_k^H\bf{W}_{p,m}$ is independent from the term $g_{mk^\prime}$ similar to \cite{marzetta_free16}, \textit{Appendix A}, the term $A$ in (\ref{eiui}) immediately is given by
$
A = N q_{k^\prime} \sum_{m=1}^Mc_{mk}^2u_{mk}^2\beta_{mk^\prime}.
$
The term $B$ in (\ref{eiui}) can be obtained as
\vspace{-.1in}
\begin{IEEEeqnarray}{rCl}
 \label{bb111} 
&&B
=
  \underbrace{\tau_p p_p q_{k^\prime}\mathbb{E}\left \{\left |\sum_{m=1}^M c_{mk}u_{mk} ||{\mathbf{g}}_{mk^\prime}||^2\pmb{\phi}_k^H{\pmb{\phi}}_{k^\prime}\right |^2\right \}}_{C}
\nonumber
\\
&+&
\underbrace{\tau_p p_p q_{k^\prime}\mathbb{E} \left\{\left | \sum_{m=1}^M c_{mk}u_{mk} \left(\sum_{i\ne k^\prime}^{K}\mathbf{g}_{mi}\pmb{\phi}_k^H\pmb{\phi}_i
\right)^H {\mathbf{g}}_{mk^\prime}\right |^2 \right\}}_{D}\!.
\end{IEEEeqnarray}
The first term in (\ref{bb111}) is given by
\vspace{-.1in}
\begin{IEEEeqnarray}{rCl}
C &=& \tau_p p_p q_{k^\prime}\mathbb{E}\left \{\left |\sum_{m=1}^M c_{mk}u_{mk}||{\mathbf{g}}_{mk^\prime}||^2\pmb{\phi}_k^H{\pmb{\phi}}_{k^\prime} \right| ^2 \right\} \nonumber
 \\
&=&
N \tau_p p_p q_{k^\prime}\left |\pmb{\phi}_k^H{\pmb{\phi}}_{k^\prime}\right |^2\sum_{m=1}^Mc_{mk}^2u_{mk}^2\beta_{mk^\prime}^2
\nonumber
\\
&+&
 N^2q_{k^\prime}\left |\pmb{\phi}_k^H{\pmb{\phi}}_{k^\prime}\right |^2\left(\sum_{m=1}^M u_{mk}\gamma_{mk}\dfrac{\beta_{mk^\prime}}{\beta_{mk}}\right)^2,
 \label{CC}
\end{IEEEeqnarray}where the last equality is derived based on the fact $\gamma_{mk}=\sqrt{\tau_p p_p}\beta_{mk}c_{mk}$. The second term in (\ref{bb111}) can be obtained as
\vspace{-.08in}
\begin{IEEEeqnarray}{rCl}
D
&=&
 \tau_p p_p q_{k^\prime}\mathbb{E}\left \{\left |\sum_{m=1}^M c_{mk}u_{mk}\Big(\sum_{i\ne k^\prime}^{K}\mathbf{g}_{mi}\pmb{\phi}_k^H\pmb{\phi}_i\Big)^H{\mathbf{g}}_{mk^\prime}\right |^2 \right\}
\nonumber
 \\
&=\!&\! N\sqrt{\tau_p p_p}q_{k^\prime}\sum_{m=1}^{M}\!u_{mk}^2c_{mk}\beta_{mk^\prime}\beta_{mk}
\!-\!
Nq_{k^\prime}\!\sum_{m=1}^{M}u_{mk}^2c_{mk}^2\beta_{mk^\prime}
\nonumber
\\
&-&
N\tau_p p_p q_{k^\prime}\sum_{m=1}^{M}u_{mk}^2c_{mk}^2\beta_{mk^\prime}^2\left| \pmb{\phi}_k^H{\pmb{\phi}}_{k^\prime}\right|^2.
 \label{d}
\end{IEEEeqnarray}
Finally by substituting (\ref{CC}) and (\ref{d}) into (\ref{bb111}), and substituting (\ref{bb111}) into (\ref{eiui}), we obtain
\vspace{-.06in}
\begin{IEEEeqnarray}{rCl}
 \label{euiu_final}
\mathbb{E}\{| \text{IUI}_{k k^\prime}|^2\} 
&=&
N\rho q_{k^\prime}\left(\sum_{m=1}^{M}u_{mk}^2\beta_{mk^\prime}\gamma_{mk}\right)
\nonumber
\\
&+&
 N^2 \rho q_{k^\prime}\left|\pmb{\phi}_k^H{\pmb{\phi}}_{k^\prime}\right|^2 \left(\sum_{m=1}^{M}u_{mk} \gamma_{mk}\dfrac{\beta_{mk^\prime}}{\beta_{mk}}\right)^2.
\end{IEEEeqnarray}
The total noise for the user $k$ is given by
\vspace{-.03in}
\begin{IEEEeqnarray}{rCl}
\!\!\!\mathbb{E}\left\{\left|\text{TN}_k\right|^2\right\}
=
\mathbb{E}\left\{\left|\sum_{m=1}^{M}u_{mk}\hat{\mathbf{g}}_{mk}^H\mathbf{n}_m\right|^2\right\}
=
N\sum_{m=1}^{M}u_{mk}^2\gamma_{mk},
\label{tn}
\vspace{-.05in}
\end{IEEEeqnarray}where the last equality is due to the fact that the terms $\hat{\mathbf{g}}_{mk}$ and $\mathbf{n}_m$ are uncorrelated.
Based on the analysis in \cite{mezghani_WSA16}, we have
\vspace{-.06cm}
\begin{IEEEeqnarray}{rCl}
 \mathbf{R}_{e_k^ze_k^z}
 \stackrel{(a)}{=}\sigma_{\dot{e}}^2~\text{diag}(\mathbf{R}_{z_kz_k}),
 \end{IEEEeqnarray}where $\mathbf{R}_{e_k^ze_k^z}$ and $\mathbf{R}_{z_kz_k}$ refer to the covariance matrix of the quantization error and the covariance matrix of the input of the  quantizer, respectively. Moreover, note that in step (a), we exploit the analysis in  \cite[Section V]{mezghani_WSA16}. Thus, the power of the quantization error for user $k$ is given by:
\begin{IEEEeqnarray}{rCl}
\!\!\!\!\!\mathbb{E}\left\{\left|\text{TQE}_k\right|^2\right\}=\mathbb{E}\left\{\left|\sum_{m=1}^{M}u_{mk}e_{mk}^z\right|^2\right\}=\sum_{m=1}^{M}u_{mk}^2\mathbb{E}\left\{\left|e_{mk}^z\right|^2\right\},~~
\label{tq}
\end{IEEEeqnarray}
Finally, the power of the quantization error is obtained as the following:
\begin{IEEEeqnarray}{rCl}
\mathbb{E}\left\{\left|e_{mk}^z\right|^2\!\right\}=\mathbb{E}\left\{\left|\dot{e}_{mk}^z\right|^2\!\right\} \sigma_{z_{mk}}^2=\sigma_{\dot{e}}^2 \sigma_{z_{mk}}^2,
\end{IEEEeqnarray}
where we used the fact that all APs use the same number of bits to quantize the weighted signal $z_{mk}$ in (\ref{rkcase3}). Next, the term $\sigma_{z_{mk}}^2$ is obtained as
\begin{IEEEeqnarray}{rCl}\label{e1}
&&\!\!\!\!\sigma_{z_{mk}}^2\!=\! \mathbb{E}\!\left\{ \left(\hat{\mathbf{g}}_{mk}^{H}\mathbf{y}_m\right)^H\left(\hat{\mathbf{g}}_{mk}^{H}\mathbf{y}_m\right)\right\} 
\!\!=\!\!
 \mathbb{E}
\Bigg\{\! \Bigg(\!    
\sqrt{\rho}\!\sum_{k^{\prime}=1}^{K}\!u_{mk}\hat{\mathbf{g}}_{mk}^H\mathbf{g}_{mk^\prime}\sqrt{q_{k^\prime}}s_{k^\prime}
\nonumber
\\
&+&u_{mk}\hat{\mathbf{g}}_{mk}^H\mathbf{n}_m
 \Bigg)^H
 \!\left(\!\sqrt{\rho}\!\sum_{k^{\prime}=1}^{K}\!\!u_{mk}\hat{\mathbf{g}}_{mk}^H\mathbf{g}_{mk^\prime}\sqrt{q_{k^\prime}}s_{k^\prime}
 \!\!+\!\!
 u_{mk}\hat{\mathbf{g}}_{mk}^H\mathbf{n}_m
  \!\right)\!\!\Bigg\}
  \nonumber
  \\
&\!\approx\!&
 \rho\mathbb{E}\left\{\left|\sum_{k^{\prime}=1}^{K}u_{mk}\hat{\mathbf{g}}_{mk}^H\mathbf{g}_{mk^\prime}\sqrt{q_{k^\prime}}s_{k^\prime}\right|^2\right\}
 \!+\!
 \mathbb{E}\left\{\left|u_{mk}\hat{\mathbf{g}}_{mk}^H\mathbf{n}_m \right|^2\right\}\!,\!\!\!
\end{IEEEeqnarray}
where the approximation (\ref{e1}) is obtained byignoring the correlation between the terms $\hat{\mathbf{g}}_{mk}^H\mathbf{n}_m$ and $\sum_{k^{\prime}=1}^{K}\hat{\mathbf{g}}_{mk}^H\mathbf{g}_{mk^\prime}\sqrt{q_{k^\prime}}s_{k^\prime}$. Note that the simulation results confirm that 
this approximation is very tight \cite{our_tgcn_accepted}.
The first term in (\ref{e1}) can be obtained as

\begin{small}
\begin{IEEEeqnarray}{rCl}
&&\mathbb{E}\left\{ \left| \sum_{k^{\prime}=1}^{K}u_{mk}\hat{\mathbf{g}}_{mk}^H\mathbf{g}_{mk^\prime}\sqrt{q_{k^\prime}}s_{k^\prime} \right|^2 \right\}\nonumber
\\
 &=&
 \mathbb{E}\left\{ \left| \sum_{k^{\prime}=1}^{K}u_{mk}(\mathbf{g}_{mk}-\bep_{mk})^H\mathbf{g}_{mk^\prime}\sqrt{q_{k^\prime}}s_{k^\prime} \right|^2 \right\}
\nonumber
\\
 & \! \!=&
 \! \!\underbrace{\mathbb{E} \!\left\{\left|
 \sum_{k^{\prime}=1}^{K} \!u_{mk}{\mathbf{g}}_{mk}^H\mathbf{g}_{mk^\prime}\sqrt{q_{k^\prime}}s_{k^\prime} \right|^2\right\}}_{\text{I}}
  \!+ \!
 \underbrace{\mathbb{E}\left\{\left|\sum_{k^{\prime}=1}^{K} \!u_{mk}{\bep}_{mk}^H\mathbf{g}_{mk^\prime}\sqrt{q_{k^\prime}}s_{k^\prime}\right|^2\!\right\}}_{\text{II}},~~~~~
\label{e3}
\end{IEEEeqnarray}
\end{small}where each element of $\bep$ is given by $\epsilon_{mk}= \mathcal{CN}(0,\beta_{mk}-\gamma_{mk})$. The terms I and II in (\ref{e3}) are given as following:
$
\text{I} = N \beta_{mk}\sum_{k^\prime=1}^{K}q_{k^\prime}\beta_{mk^\prime},
$
and
$
\text{II} = N (\beta_{mk}-\gamma_{mk})\sum_{k^\prime=1}^{K}q_{k^\prime}\beta_{mk^\prime}.
$
Finally, we have
\begin{IEEEeqnarray}{rCl}
&&\mathbb{E}\left\{\left|\text{TQE}_k\!\right|^2\right\}
\nonumber
\\
&\approx&
N \sigma_{\dot{e}}^2\sum_{m=1}^Mu_{mk}^2 \left[ \sqrt{\rho}\left(2\beta_{mk}-\gamma_{mk}\right)\sum_{k^\prime=1}^{K}q_{k^\prime}\beta_{mk^\prime} + \gamma_{mk}\right],~~
\label{e6}
\vspace{-.2cm}
\end{IEEEeqnarray}
By substituting (\ref{dsk_vector}), (\ref{ebuk}), (\ref{euiu_final}) and (\ref{tn}) into (\ref{sinrdef11}), the corresponding SINR of the $k$th user is obtained by (\ref{sinrcpustat}), which completes the proof of Theorem \ref{theorem_up_quan_case3}.
~~~~~~~~~~~~~~~~~~~~~~~~~~~~~~~$\blacksquare$
\section*{Appendix D: Proof of Proposition \ref{prop_p5_standard}}
The standard form of GP is defined as follows \cite{bookboyd}: 
\begin{subequations}
\label{p12} 
\begin{align}P_{12}:~~~~
\label{p12_1}\min \quad & f_0(\mathbf{x}),\\
\label{p12_2}\!\!\!\!\!\!\!\!\!\!\!\!\!\!\!\!\text{subject to}\quad &f_i(\mathbf{x}) \le 1, i= 1,\cdots, m, g_i(\mathbf{x}) = 1, i= 1,\cdots, p,
\end{align}
\end{subequations}
where $f_0$ and $f_i$ are posynomial and $g_i$ are monomial functions. Moreover, $\mathbf{x}=\{x_1,\cdots,x_n\}$ represent the optimization variables. The SINR constraint in (\ref{p12}) is not a posynomial function in this form, however it can be rewritten as the following posynomial function:
\begin{IEEEeqnarray}{rCl}
\small
&&\!\!\!\!\!\!
\dfrac{\mathbf{u}_k^H\!\left(\!N^2\!\sum_{k^\prime\ne k}^K\!q_{k^\prime}|\pmb{\phi}_k^H\pmb{\phi}_{k^\prime}|^2\blam_{k k^\prime}
\blam_{k k^\prime}^H
\!+\!
N\sum_{k^\prime=1}^{K}q_{k^\prime}\bup_{kk^\prime}
\!+\!\!
\dfrac{N\mathbf{R}_{k}}{\rho}\right)\mathbf{u}_k}{\mathbf{u}_k^H\left(N^2q_k\bgama_k\bgama_k^H\!\right)\!\!\mathbf{u}_k}
\nonumber
\\
&<&
\dfrac{1}{t},
 \forall k.
\label{inv}
\end{IEEEeqnarray}
By applying a simple transformation, (\ref{inv}) is equivalent to the following inequality:
\begin{IEEEeqnarray}{rCl}
q_k^{-1}\left(\sum_{k^\prime\ne k}^K\!a_{kk^\prime}q_{k^\prime}
+
\sum_{k^\prime=1}^{K}b_{kk^\prime}q_{k^\prime}+c_k\right)<\dfrac{1}{t}, \forall k,
\label{inv2}
\end{IEEEeqnarray}
where 
$a_{kk^\prime}=\frac{\mathbf{u}_k^H \left(\left|\!\pmb{\phi}_k^H\!\pmb{\phi}_{k^\prime}\right|^2\!\blam_{k k^\prime}\!\blam_{k k^\prime}^H\right) \mathbf{u}_k}{\mathbf{u}_k^H\!\left(\bgama_k\bgama_k^H\right)\mathbf{u}_k}$, $b_{kk^\prime}=\frac{\mathbf{u}_k^H \bup_{k k^\prime}\mathbf{u}_k}{\mathbf{u}_k^H\left(\bgama_k\bgama_k^H\right)\mathbf{u}_k}$ and
$c_k=\frac{\mathbf{u}_k^H \mathbf{R}_{k} \mathbf{u}_k}{\rho\mathbf{u}_k^H\left(\bgama_k\bgama_k^H\right)\mathbf{u}_k}$.
The transformation in (\ref{inv2}) shows that the left-hand side of (\ref{inv}) is a posynomial function. Hence, the power allocation Problem $P_4$ is a GP (convex problem), where the objective function and constraints are monomial and posynomial, respectively, which completes the proof of Proposition \ref{prop_p5_standard}.~~~
~~~~~~~~~~~~~~~~~~~~~~~~~~~~~~~~~~~~~~~~~~~~~~~~~~~~$\blacksquare$
\section*{Appendix E: Proof of Lemma \ref{lemma_p2_p6}}
This lemma is proven by exploiting the unique optimal solution of the uplink max-min SINR problem with total power limitation through an eigensystem \cite{Bochetvt4}. This problem is iteratively solved and the optimal receiver filter coefficients $\check{\mathbf{U}}$ are determined by solving Problem $P_3$.
Next, we scale the power allocation at each user such that the per-user power constraints are satisfied. 
Let us consider the following optimization problem for a given receiver filter coefficients $\check{\mathbf{U}}$:
\vspace{-.02cm}
\begin{subequations}
\label{plemma2q} 
\begin{align}\!\!\!\!\!\!\!\!\!P_{11}:
\label{plemma2q_1}C^{\text{UP}}\left(\check{\mathbf{U}},P_{\text{tot}}\right)=\max_{q_k}~~~~ &\min_{k=1,\cdots,K}\quad  \text{SINR}_k^{\text{UP}}\left(\check{\mathbf{U}},\mathbf{q}\right),
\\\vspace{-.3cm}
\label{plemma2_2}~~~~~~~~~~~~\text{subject to}\quad &  \sum_{k=1}^Kq_k \le P_{\text{tot}}.
\end{align}
\end{subequations}
The optimal solution of Problem $P_{11}$ can be determined by finding the unique eigenvector associated with unique positive eigenvalue of an eigensystem and the power allocation $\check{\mathbf{q}}$ that satisfies the following condition \cite{Bochetvt4}:
\begin{equation}
\sum_{k=1}^K \check{q}_k =  P_{\text{tot}}.
\end{equation}
The SINRs of all users can be collectively written as
\vspace{-.07cm}
\begin{equation}
\label{kequations}
\check{\mathbf{q}}\frac{1}{C_{k}^{\text{UP}}\left(\check{\mathbf{U}},P_{\text{tot}}\right)}=\mathbf{D}\bpsi\left(\check{\mathbf{U}}\right)\check{\mathbf{q}}+\mathbf{D}\bsigma\left(\check{\mathbf{U}}\right),
\end{equation}
\vspace{-.17cm}
where $\bsigma\left(\check{\mathbf{U}}\right)\in \mathbb{C}^{K\times 1}$, $\sigma_k\left({\mathbf{u}}_k\right)=\dfrac{\!N}{\!\rho}\left(\!\!\dfrac{\sigma_{\dot{e}}^2}{\textcolor{black}{\dot{a}}^2}\!+\!1\right)\!\sum\limits_{m=1}^{M}\!\check{u}_{mk}\gamma_{mk}$ and $\mathbf{D}$ and $\bpsi\left(\check{\mathbf{U}}\right)$ are defined as
\begin{equation}
\mathbf{D}=\text{diag}\left[\dfrac{1}{\check{\mathbf{u}}_1^H\check{\mathbf{D}}_1\check{\mathbf{u}}_1},\cdots,\dfrac{1}{\check{\mathbf{u}}_K^H\check{\mathbf{D}}_K\check{\mathbf{u}}_K}\right],
\end{equation}
\begin{equation}
{
 \left[\bpsi\left(\check{\mathbf{U}}\right)\right]_{kk^\prime}
 =
 \left\{
\!\!\!\!\!\!\!
\begin{array}{rl}
&\check{\mathbf{u}}_k^H\check{\check{\mathbf{R}}}_{kk}\check{\mathbf{u}}_k,
~~~~~~~~~~~~~~k=k^\prime,\\
&\check{\mathbf{u}}_k^H\check{\mathbf{R}}_{kk^\prime}\check{\mathbf{u}}_k+\check{\mathbf{u}}_k^H\check{\check{\mathbf{R}}}_{kk^\prime}\check{\mathbf{u}}_k ,k\neq k^\prime ,
\end{array}
 \right.}
\end{equation}where ${\check{\mathbf{D}}_k}$, $\check{\mathbf{R}}_{kk^\prime}$ and $\check{\check{\mathbf{R}}}_{kk^\prime}$ are defined as
\begin{IEEEeqnarray}{rCl}
\label{sinr_lemma2}
&&\text{SINR}_k^{\text{UP}}  =
\\
&& \dfrac{q_k\mathbf{u}_k^H\Big(\overbrace{N^2\bgama_k\bgama_k^H}^{\check{\mathbf{D}}_k}\Big)\mathbf{u}_k}
{\mathbf{u}_k^H\!\Big(\!\sum\limits_{k^\prime\ne k}^K\!q_{k^\prime}\underbrace{N^2|\pmb{\phi}_k^H\pmb{\phi}_{k^\prime}|^2\blam_{k k^\prime}\blam_{k k^\prime}^H}_{\check{\mathbf{R}}_{kk^\prime}}\!+\!
\sum\limits_{k^\prime=1}^{K}\!q_{k^\prime}\underbrace{N\bup_{kk^\prime}}_{\check{\check{\mathbf{R}}}_{kk^\prime}}+\dfrac{N\mathbf{R}_{k}}{\rho}\!\Big )\mathbf{u}_k}\!.\nonumber
\end{IEEEeqnarray}
Having both sides of (\ref{kequations}) multiplied by $\textbf{1}^T=\left[1,\cdots,1\right]$, we obtain
$
\frac{1}{C_{k}^{\text{UP}}\left(\check{\mathbf{U}},P_{\text{tot}}\right)}\!=\!\dfrac{1}{P_{\text{tot}}}\textbf{1}^T\check{\mathbf{D}}\bpsi\left(\check{\mathbf{U}}\right)\check{\mathbf{q}}+\dfrac{1}{P_{\text{tot}}}\textbf{1}^T\mathbf{D}\bsigma\left(\check{\mathbf{U}}\right)
$, which can be combined with (\ref{kequations}) to define the following eigensystem:
\begin{equation}
\small \blam\left(\check{\mathbf{U}},P_{\text{tot}}\right)\check{\mathbf{q}}_{\text{ext}}=\frac{1}{C_{k}^{\text{UP}}\left(\check{\mathbf{U}},P_{\text{tot}}\right)}\check{\mathbf{q}}_{\text{ext}},
\end{equation}
\begin{equation}
\label{eigenssystem}
\blam\left(\check{\mathbf{U}},P_{\text{tot}}\right)=\begin{bmatrix}
\mathbf{D}\bpsi^T\left(\check{\mathbf{U}}\right) & \mathbf{D}\bsigma\left(\check{\mathbf{U}}\right) \\
\dfrac{1}{P_{\text{tot}}}\textbf{1}^T \mathbf{D}\bpsi^T\left(\check{\mathbf{U}}\right) & \dfrac{1}{P_{\text{tot}}}\textbf{1}^T \mathbf{D}\bsigma\left(\check{\mathbf{U}}\right)
\end{bmatrix}\!\!.\!\!
\end{equation}
The optimal power allocation $\check{\mathbf{q}}$ is obtained by determining the eigenvector corresponding to the maximum eigenvalue of $\blam\left(\check{\mathbf{U}},P_{\text{tot}}\right)$ and scaling the last element to one as follows:
\begin{equation}
\blam\left(\check{\mathbf{U}},P_{\text{tot}}\right)\check{\mathbf{q}}_{\text{ext}}=\lambda_{\text{max}} \left(\blam\left(\check{\mathbf{U}},P_{\text{tot}}\right)\right)\check{\mathbf{q}}_{\text{ext}},~\check{\mathbf{q}}_{\text{ext}}=\begin{bmatrix} \check{\mathbf{q}}\\1\end{bmatrix},
\label{optimalq}
\end{equation}
Note that to find the optimal power allocation $\check{\mathbf{q}}$, the elements of eigenvector of $\blam\left(\check{\mathbf{U}},P_{\text{tot}}\right)$ should be scaled such that the last element is one to satisfy the total power constraint. In particular, the element of the eigenvector that needs to be scaled depends on the type of power constraint in the problem. For example, to meet the total power constraint, the last element is scaled to one. Similarly, to meet the other types of power constraints (for example, per-user power constraint), the components of this eigenvector can be scaled by any positive value to satisfy a given  condition as follows:
\begin{equation}
\label{problem_scale_both}
\blam\left(\check{\mathbf{U}},P_{\text{tot}}\right)\delta_{\text{cons}}\check{\mathbf{q}}_{\text{ext}}
=
\lambda_{\text{max}} \left(\blam\left(\check{\mathbf{U}},P_{\text{tot}}\right)\right)\delta_{\text{cons}}\check{\mathbf{q}}_{\text{ext}},
\vspace{-.02cm}
\end{equation}where $\delta_{\text{cons}}$ is a positive constant. This is the key fact that exploited to show that both Problems $P_2$ and $P_6$ provide the same optimal solution. We further scale the power allocation $\check{\mathbf{q}}$ to satisfy the per-user power constraints which is performed through carrying out the following two steps:
\begin{equation}
\small
\label{hatq}
\bar{\mathbf{q}}= 
\begin{bmatrix} 
\dfrac{\check{q}_1}{p_{\text{max}}^{(1)}}\\\vdots\\\dfrac{\check{q}_K}{p_{\text{max}}^{(K)}}.
\end{bmatrix}.
\end{equation}
Next, we find the maximum value among the elements of $\bar{\mathbf{q}}$, i.e., $\max(\bar{\mathbf{q}})$, and divide all elements of $\bar{\mathbf{q}}$ by it. Hence the power allocation $\check{\check{\mathbf{q}}}$ is defined as follows:
\begin{equation}
\small
\label{hathatq}
\check{\check{\mathbf{q}}}=\begin{bmatrix} 
\dfrac{\check{q}_1}{\max(\bar{\mathbf{q}})}\\\vdots\\\dfrac{\check{q}_K}{\max(\bar{\mathbf{q}})},
\end{bmatrix}.
\end{equation}
In the next iteration, the same max-min problem is solved with a new total power constraint obtained by summing up the allocated power to all users in the previous iteration:
\begin{subequations}
\label{plemma6qeq} 
\begin{align}\!\!L_1\!:
\label{plemma6qeq_1}\!C^{\text{UP}}\!\left(\check{\mathbf{U}},P_{\text{tot}}^{\text{new}}\right)\!=\!\max_{q_k}~ \min_{k=1,\cdots,K}  \text{SINR}_k^{\text{UP}}\left(\check{\mathbf{U}},\mathbf{q}\right) ,\\ \vspace{-.12cm}
\label{plemma6qeq_2}\text{subject to}~~~   \sum\limits_{k=1}^K q_k \le P_{\text{tot}}^{\text{new}},~~\text{where}~~ P_{\text{tot}}^{\text{new}}=\sum\limits_{k=1}^K\check{\check{q}}_k.
\end{align}
\end{subequations}
At the convergence of the algorithm, the per-user power constraints are satisfied with achieving the same uplink SINR for each user. Interestingly, if this max-min problem is solved with the corresponding total power constraint, then it will converge to the same optimal solution of max-min problem with per-user power constraints. This is due to the property that the eigensystem exploited to obtain the power allocation in (\ref{eigenssystem}) has a unique positive eigenvalue and a corresponding unique eigenvector. 
Furthermore, in both Problems $P_2$ and $P_6$, different elements of the same eigenvector are scaled to meet the corresponding constraints on the power allocation. In other words, the last element is scaled to meet the total power constraint in $P_6$ whereas the element with the highest ratio as in (\ref{problem_scale_both}) is scaled to meet the per-user power constraint. 
As the equivalent total power $P_{\text{tot}}^c$ for Problem $P_6$ chosen from the solution of the original $P_2$, both of them will converge to the same solution whose optimality is proven later by considering an equivalent  problem related to the virtual downlink SINR.
Therefore, Problems $P_2$ and $P_6$ are equivalent and have the same optimal solution. 
~~~~~~~~~~~~~~~~~~~~~~~~~~~~~~~~~~~~~~~~~~~~~~~~~~~~
~~~~$\blacksquare$
\section*{Appendix F: Proof of Theorem \ref{theorem_quan_updl_power}} 
To achieve the same SINR tuples in both the uplink and the downlink, we need:
\begin{IEEEeqnarray}{rCl}
\text{SINR}_k^{\text{DL}}\left(\mathbf{U},\mathbf{p}\right) =\text{SINR}_k^{\text{UP}}\left(\mathbf{U},\mathbf{q}\right), \forall k.
\label{dual2}
\end{IEEEeqnarray}
By substituting uplink and downlink SINRs, in (\ref{sinrupquan}) and (\ref{sinrdlquan}), respectively, in equation (\ref{dual2}) and summing all equations by both sides, we have
\begin{IEEEeqnarray}{rCl}
&&\!\!\!\!\!\!\!\!\!p_1N\sum\limits_{m=1}^M\!\left(\frac{\sigma_{\dot{e}}^2}{\textcolor{black}{\dot{a}}^2}\!+\!1\right)u_{m1}^2\gamma_{m1}\!+\!\cdots\!+\! p_KN\sum\limits_{m=1}^{M}\left(\frac{\sigma_{\dot{e}}^2}{\textcolor{black}{\dot{a}}^2}\!+\!1\right)\!u_{mK}^2\gamma_{mK}
\nonumber
\\
&=&
\sum\limits_{k=1}^Kq_{k}.
\label{dual3}
\end{IEEEeqnarray}
Therefore, this condition between the total transmit power on the uplink and the equivalent total transmit power on the downlink should be satisfied to realize the same SINRs for all users.~~~~~~~~~~~~~~~~~~~~~~~~
~~~~~~~~~~~~~~~~~~~~~~~~~~~~~~~~~~~~~~~~~~~~~~~~~~~~~~~~
$\blacksquare$
\bibliographystyle{IEEEtran}
\bibliography{final1_cuma} 
\end{document}